\documentclass[11pt]{article}\pdfoutput=1
\usepackage{cite}
\usepackage{amsmath,amsfonts,amssymb}
\usepackage[small,bf,hang]{caption}
\usepackage{slashed}
\usepackage{mathabx,amsmath}
\usepackage{latexsym,epsfig}
\usepackage{arydshln}

\def\hybrid{
        \topmargin -20pt
        \oddsidemargin 0pt
        \headheight 0pt \headsep 0pt
        \textwidth 6.25in 
        \textheight 9.5in 
        \marginparwidth .875in
        \parskip 5pt plus 1pt \jot = 1.5ex}

\hybrid

 \csname
@addtoreset\endcsname{equation}{section}

\def\moth{\mathsurround=0pt}
\newdimen\zo \zo=0pt

\def\tick{\leaders\hrule height 0.5ex depth 0pt \hskip 0.5pt}
\def\upboxfill{$\moth \setbox\zo\hbox{\tick}%
  \hskip 3pt\hbox to 0pt{$\tick$\hss}\hrulefill \hbox to 7.5pt{$\tick$\hss}$}

\def\dtick{\leaders\hrule height .34pt depth 0.5ex \hskip 0.5pt}
\def\downboxfill{$\moth \setbox\zo\hbox{\dtick}%
  \hskip 2pt\hbox to 0pt{$\dtick$\hss}\hrulefill \hbox to 2pt{$\dtick$\hss}$}

\def\bec{\begin{center}}
\def\ec{\end{center}}

\def\d{\delta}

\def\cB{{\cal B}}

\def\cA{{\cal A}}

\def\cA{{\cal A}}

\def\nn{\nonumber}

 \def\det{{\rm det\,}}
\def\be{\begin{equation}}
\def\ee{\end{equation}}
\def\bea{\begin{eqnarray}}
\def\eea{\end{eqnarray}}
\def\ba{\begin{array}}
\def\ea{\end{array}}

\def\ft#1#2{{\textstyle{{\scriptstyle #1}
\over {\scriptstyle #2}}}}

\newcommand{\eft}{{EFT\ }}
\newcommand{\bfa}{{\boldsymbol{\alpha}}}
\newcommand{\bfb}{{\boldsymbol{\beta}}}

\begin{document}

\begin{titlepage}
\rightline{}
\rightline{\tt }
\rightline{\tt  MIT-CTP-4604}
\rightline{October 2014}
\begin{center}
\vskip 1.0cm
{\Large \bf {Consistent Kaluza-Klein Truncations \\[0.5ex] via\\[1.25ex]  Exceptional Field Theory}}\\
\vskip 1.2cm
{\large {Olaf Hohm${}^1$ and Henning Samtleben${}^2$}}
\vskip .6cm
{\it {${}^1$Center for Theoretical Physics}}\\
{\it {Massachusetts Institute of Technology}}\\
{\it {Cambridge, MA 02139, USA}}\\
ohohm@mit.edu
\vskip 0.2cm
{\it {${}^2$Universit\'e de Lyon, Laboratoire de Physique, UMR 5672, CNRS}}\\
{\it {\'Ecole Normale Sup\'erieure de Lyon}}\\
{\it {46, all\'ee d'Italie, F-69364 Lyon cedex 07, France}}\\
henning.samtleben@ens-lyon.fr

\vskip 1cm
{\bf Abstract}
\end{center}


\begin{narrower}

\noindent
We present the generalized Scherk-Schwarz reduction ansatz 
for the full supersymmetric exceptional field theory in terms of group valued twist matrices
subject to consistency equations.
With this ansatz the field equations  
precisely reduce to those of lower-dimensional gauged supergravity
parametrized by an embedding tensor.
We explicitly construct a family of twist matrices as solutions of the
consistency equations. They induce gauged supergravities with gauge groups 
SO$(p,q)$ and CSO$(p,q,r)$. 
Geometrically, they describe compactifications on internal spaces given by  spheres and 
(warped) hyperboloides $H^{p,q}$, thus extending the applicability of generalized 
Scherk-Schwarz reductions beyond homogeneous spaces.
Together with the dictionary that relates exceptional field theory to $D=11$ and IIB supergravity,
respectively, the construction defines an entire new family of consistent truncations of the 
original theories. These include not only compactifications on spheres of different dimensions 
(such as AdS$_5\times S^5$), but also various hyperboloid compactifications giving rise to 
a higher-dimensional embedding of supergravities with non-compact and 
non-semisimple gauge groups.

\end{narrower}

\end{titlepage}

\newpage
\tableofcontents

\newpage

\section{Introduction}

The consistent Kaluza-Klein truncation of higher-dimensional supergravity to 
lower-dimensional theories is an important and in general surprisingly difficult problem. 
Here consistent truncation means  that any solution of the 
lower-dimensional theory can be embedded 
into a solution of the original, higher-dimensional theory. 
This requires that all 
coordinate dependence of the internal space is consistently factored out.
Due to the non-linearity of the supergravity equations of motion this 
is a highly non-trivial and often impossible challenge for compactifications on curved backgrounds.  
Only very few examples are known in which such consistency cannot be attributed to an underlying symmetry argument.
The simplest class of consistent truncations are the $T^n$ toroidal compactifications in which the internal coordinate 
dependence is completely dropped, extrapolating the original ideas of Kaluza and Klein~\cite{Kaluza:1921tu,Klein:1926tv} 
to higher dimensions. Consistency simply follows from the fact that all retained massless fields are singlets under the resulting 
${\rm U}(1)^n$ gauge group. 
In the context of eleven-dimensional supergravity \cite{Cremmer:1978km} such reductions give rise 
to the maximal ungauged (thus abelian) supergravities in lower dimensions~\cite{Cremmer:1979up}.

More involved examples are sphere compactifications, the prime example being the compactification of
eleven-dimensional supergravity on AdS$_4\times {\rm S}^7$, 
leading to maximal ${\rm SO}(8)$ gauged supergravity in four 
dimensions \cite{deWit:1982ig}. The required consistency conditions 
are so non-trivial that in the early days of Kaluza-Klein supergravity this shed serious doubt
on the possible consistency of sphere compactifications.
For AdS$_4\times {\rm S}^7$ this consistency was, however, established in \cite{deWit:1986iy}, with recent improvements 
in \cite{Nicolai:2011cy,deWit:2013ija,Godazgar:2013dma,Godazgar:2013pfa}, employing an ${\rm SU}(8)$ invariant reformulation of the original eleven-dimensional 
theory~\cite{deWit:1986mz}. Other consistent sphere reductions have been constructed in \cite{Nastase:1999kf,Cvetic:2000dm,Cvetic:2000nc}, including the compactification on AdS$_7\times {\rm S}^4$.

An important generalization of the usual compactification scheme was put forward by Scherk and Schwarz \cite{Scherk:1979zr}, relating the internal dimensions
to the manifold of a Lie group.
More recently, the advances in the understanding of the duality symmetries underlying string and M-theory
 have nourished the idea
to identify generalized geometric (and possibly non-geometric) compactifications as generalized Scherk-Schwarz reductions in some 
extended geometry~\cite{Kaloper:1999yr,Hull:2004in,Dabholkar:2005ve,Hull:2007jy,DallAgata:2007sr,Hull:2009sg}.
In the framework of 
double field theory~\cite{Siegel:1993th,Hull:2009mi,Hull:2009zb,Hohm:2010jy,Hohm:2010pp,Hohm:2010xe}, 
which makes
the ${\rm O}(d,d)$ T-duality of string theory manifest, 
generalized Scherk-Schwarz-type compactifications of 
an extended spacetime have been 
discussed in~\cite{Hohm:2011cp,Aldazabal:2011nj,Geissbuhler:2011mx,Hassler:2014sba}, 
see also \cite{Grana:2012rr,Hohm:2011ex} for reductions to deformations of double field theory. 
In the M-theory case, analogous ideas have been investigated in 
\cite{Berman:2012uy,Musaev:2013rq,Aldazabal:2013mya,Berman:2013cli,Lee:2014mla,Baron:2014yua}
in the duality covariant formulation of the internal sector of 
$D=11$ supergravity~\cite{Berman:2010is,Berman:2011jh,Coimbra:2011ky,Coimbra:2012af}.

In this paper, we realize this scenario in full exceptional field theory (EFT) 
\cite{Hohm:2013pua,Hohm:2013vpa,Hohm:2013uia,Hohm:2014fxa,Godazgar:2014nqa},
which is the manifestly U-duality covariant formulation of the untruncated ten- and eleven-dimensional supergravities. 
The theory is formulated on a generalized spacetime coordinatized by $(x^{\mu},Y^M)$, where 
we refer to $x^{\mu}$ as `external' spacetime coordinates, while the $Y^M$ describe some generalized  
`internal' space with $M,N$ labeling the fundamental representation of the Lie groups in the 
exceptional series E$_{d(d)}$, $2\leq d\leq 8$. 
The fields generically include an external metric $g_{\mu\nu}$, 
an internal (generalized) metric ${\cal M}_{MN}$ and various higher $p$-forms, in particular 
Kaluza-Klein-like vectors ${\cal A}_{\mu}{}^{M}$ in the fundamental representation and possibly 
2-forms and higher forms.
With respect to the internal space, all fields are
subject to covariant section constraints on the extended derivatives $\partial_M$ which imply that 
fields depend only on a subset of coordinates. There are at least two inequivalent solutions to these constraints:
for one the theory is on-shell equivalent to 11-dimensional supergravity, for the other to type IIB,
in analogy to type II double field theory \cite{Hohm:2011zr}.

The recent ideas of realizing non-trivial (and possibly non-geometric) compactifications as generalized 
Scherk-Schwarz compactifications are based on an ansatz for the generalized metric of the form 
\bea
{\cal M}_{MN}(x,Y) &=& U_{M}{}^{K}(Y)\,U_{N}{}^{L}(Y)\,M_{KL}(x)\;, 
\label{genSS_intro}
\eea
in terms of group-valued twist matrices $U_M{}^N$ which capture the $Y$-dependence of the fields. 
With this ansatz, the $Y$-dependence in the corresponding part of the field equations 
consistently factors out, provided the twist matrices satisfy a particular set of first order differential 
equations which in full generality take the form
\bea
\left[(U^{-1})_M{}^P (U^{-1})_N{}^L \, \partial_P U_L{}^K\right]_{(\mathbb{P})} &\stackrel{!}{=}&
\rho \,\Theta_M{}^\bfa\, (t_\bfa)_N{}^K
\;,
\nonumber\\
  \partial_N (U^{-1})_M{}^N  
- (D-1)\,\rho^{-1}\partial_N \rho  \, (U^{-1})_M{}^N  &\stackrel{!}{=}& 
\rho\,(D-2)\,\vartheta_M
\;.
\label{consistent_intro}
\eea
Here, $\rho$ is a $Y$-dependent scale factor, $D$ is the number of external space-time dimensions,
$\vartheta_M$ and $\Theta_M{}^\bfa$ are constant so-called embedding tensors that encode 
the gauging of supergravity, and $[\cdot]_{(\mathbb{P})}$
denotes projection onto a particular subrepresentation.
With this ansatz,
the scalar action for ${\cal M}_{MN}(x,Y)$
reproduces the scalar potential of gauged supergravity for $M_{KL}(x)$, with the twist matrix $U$
encoding the embedding tensor $\Theta_M{}^\bfa$ which parametrizes the lower-dimensional
theory \cite{Berman:2012uy,Musaev:2013rq,Aldazabal:2013mya}.

In this paper, we extend this scheme to the full exceptional field theory 
with the following main results

 \begin{itemize}
 \item[(1)] 
 We extend the ansatz (\ref{genSS_intro}) to the field content of the full exceptional field theory,
 i.e.\ to the external metric, vector and $p$-forms. In particular, we find that consistency of the
 reduction ansatz requires a particular form of the `covariantly constrained' compensating gauge fields, 
 which are novel fields required in exceptional field theory for a proper description of the degrees of freedom
 dual to those of the higher-dimensional metric.
 E.g.\ in ${\rm E}_{7(7)}$ exceptional field theory, most of 
 the remaining fields reduce covariantly, 
  \bea\label{SSFORMS}
 g_{\mu\nu}(x,Y) &=& \rho^{-2}(Y)\,g_{\mu\nu}(x)\;,\nonumber\\
  {\cal A}_{\mu}{}^{M}(x,Y) &=& \rho^{-1}(Y) A_{\mu}{}^{N}(x)(U^{-1})_{N}{}^{M}(Y) \;, 
  \nonumber\\
  {\cal B}_{\mu\nu\,\bfa}(x,Y) &=& \,\rho^{-2}(Y) U_\bfa{}^\bfb(Y)\,B_{\mu\nu\,\bfb}(x)
  \;,
 \eea
with the twist matrix $U$ in the corresponding ${\rm E}_{7(7)}$ representation and the scale factor $\rho$
taking care of the weight under generalized diffeomorphisms. 
In contrast, the constrained compensator field which in the E$_{7(7)}$ case corresponds to a 2-form ${\cal B}_{\mu\nu \,M}$ 
in the fundamental representation is subject to a non-standard Scherk-Schwarz  ansatz that reads 
\bea
{\cal B}_{\mu\nu\,M}(x,Y) ~=~
-2\, \rho^{-2}(Y)\,(U^{-1})_S{}^P(Y) \,\partial_M U_P{}^R(Y) (t^{\bfa}){}_R{}^S\, B_{\mu\nu\,\bfa}(x) 
\;,
\label{BM_intro}
\eea
relating this field to the 2-forms $B_{\mu\nu\,\bfa}$  present in gauged supergravity. 
The ansatz (\ref{SSFORMS}) for $A_{\mu}{}^{M}(x,Y)$ 
encodes the embedding of all four-dimensional vector fields and their magnetic duals. As such, it
includes the recent results of \cite{Godazgar:2013pfa} for the ${\rm SO}(8)$ sphere compactification,
but remains valid for a much larger class of compactifications, in particular, for the hyperboloids which
we explicitly construct in this paper.
The reduction ansatz for the fermionic fields in the formulation of~\cite{Godazgar:2014nqa} is remarkably
simple, their $Y$-dependence is entirely captured by a suitable power of the scale factor $\rho$\,.

We show that with the ansatz (\ref{genSS_intro})--(\ref{BM_intro}), the field equations of exceptional field theory 
precisely reduce to the field equations of the lower-dimensional gauged 
supergravity. 
Via (\ref{consistent_intro}), the twist matrix $U$ encodes the embedding tensor 
$\Theta_M{}^\bfa$, $\vartheta_M$ which specifies the field equations of the lower-dimensional
gauged supergravity~\cite{deWit:2002vt,deWit:2007mt,LeDiffon:2011wt}. 
In case $\vartheta_M\neq0$, the lower-dimensional field equations include a gauging of the
trombone scaling symmetry which in particular acts as a conformal rescaling on the metric~\cite{LeDiffon:2011wt}.
These equations do not admit a lower-dimensional action. Yet, also in this case the generalized 
Scherk-Schwarz ansatz defines a consistent truncation and we reproduce in particular the exact 
scalar contributions to the lower-dimensional field equations. 
For $\vartheta_M=0$, the reduction is also consistent on the level of the action and we 
reproduce the full action of gauged supergravity defined by an embedding tensor $\Theta_M{}^\bfa$\,.

 \item[(2)] 
The consistency of the generalized Scherk-Schwarz ansatz being guaranteed 
by the differential equations (\ref{consistent_intro}),
 it remains an equally important task to actually solve these equations. For conventional Scherk-Schwarz compactifications the existence of proper twist matrices is guaranteed by Lie's second theorem, but to our knowledge there is no corresponding 
 theorem in this generalized context. In certain cases, the existence of solutions can be inferred from additional structures on the internal manifold, such as the Killing spinors underlying the original construction of~\cite{deWit:1986iy} and then \cite{Godazgar:2013dma}, or the generalized parallelizability underlying certain coset spaces, such as the round spheres~\cite{Lee:2014mla}. 
In this paper, we explicitly construct a family of twist matrices as solutions of (\ref{consistent_intro}), that via the generalized Scherk-Schwarz ansatz give rise to gauged supergravities with gauge groups 
SO$(p,q)$ and CSO$(p,q,r)$.
Geometrically, they describe compactifications on internal spaces given by (warped) hyperboloides $H^{p,q}$
(as first conjectured in~\cite{Hull:1988jw}), thus extending the applicability of generalized Scherk-Schwarz reductions
beyond homogeneous spaces.
Our construction is based on the embedding of the linear group
${\rm SL}(n)$ into the \eft group ${\rm E}_{d(d)}$ with the internal coordinates $Y^M$ 
decomposing according to
\bea
Y^M &\longrightarrow& \left\{ Y^{[AB]},\;\dots \right\} 
\;,
\qquad\mbox{with}\quad
A, B = 0, \dots, n-1
\;,
\label{SLinE1_intro}
\eea
i.e.\ carrying the antisymmetric representation $Y^{[AB]}$.
We then construct a family of ${\rm SL}(n)$-valued twist matrices, 
parametrized by non-negative integers $(p,q,r)$ with $p+q+r=n$,
satisfying the ${\rm SL}(n)$ version of the consistency conditions (\ref{consistent_intro}). 
They depend on a subset of $n-1$ coordinates $y^i$, embedded into (\ref{SLinE1_intro})
as $y^i\equiv Y^{[0i]}$, such that the section constraint of exceptional field theory
is identically satisfied.
Upon embedding into ${\rm E}_{d(d)}$,
these twist matrices turn out to solve the full version of consistency conditions (\ref{consistent_intro}),
provided the number of external dimensions $D$ is related to $n$ as
\bea
\frac12\,(D-1) &=& \frac{n-2}{n-4}
\;.
\label{ddd_intro}
\eea
With the three principal integer solutions $(D,n)=(7,5), (5,6), (4,8)$, we thus obtain solutions
of  the consistency conditions (\ref{consistent_intro}) within ${\rm SL}(5)$, ${\rm E}_{6(6)}$, and ${\rm E}_{7(7)}$ EFT.
Their coordinate dependence is such that the reduction ansatz explicitly satisfies the \eft section constraints.
   
 \end{itemize} 

Combining these explicit solutions to the consistency equations (\ref{consistent_intro})
with the generalized Scherk-Schwark ansatz (\ref{genSS_intro})--(\ref{BM_intro}),
we thus define consistent truncations of the full exceptional field theory to
lower-dimensional supergravities with gauge groups ${\rm SO}(p,q)$ and ${\rm CSO}(p,q,r)$.
Together with the dictionary that relates exceptional field theory to $D=11$ and IIB supergravity,
respectively, (which is independent of the particular choice of the twist matrix $U$), the construction
thus gives rise to an entire family of consistent truncations in the original theories, including spheres
of various dimensions and warped hyperboloids.\footnote{
It depends on the embedding (\ref{SLinE1_intro}) of ${\rm SL}(n)$, if the coordinate dependence 
of the twist matrix falls into the class of eleven-dimensional (IIA) or IIB solutions of the exceptional field theory.
This defines in which higher-dimensional theory the construction gives rise to consistent truncations.
Unsurprisingly, this is IIA for $D=4, 7$, and IIB for $D=5$\,.
}
Specifically, we compute the internal metric induced by our twist
matrices via the Scherk-Schwarz ansatz (\ref{genSS_intro}), and find
\bea
ds^2 &=&  (1+u-v)^{-2/(p+q+r-2)} \left(
dy^z dy^z + dy^a dy^b \left(\delta_{ab} +\frac{\eta_{ac} \eta_{bd} {y}^c {y}^d}{1-v} \right)\right)
\;,
\label{internalG_intro}
\eea
with the further split of coordinates $y^i=\{y^a, y^z\}$, $a=1, \dots, p+q-1$, and $z=p+q, \dots, r$\,,
and the combinations $u\equiv y^a y^a$, $v\equiv y^a\eta_{ab}y^b$\,.
This space is conformally equivalent to the direct product
of $r$ flat directions and the hyperboloid $H^{p,q}$\,.
The three integer solutions to (\ref{ddd_intro}) in particular capture the compactifications around the 
three maximally supersymmetric solutions AdS$_7\times {\rm S}^4$, AdS$_5\times {\rm S}^5$, AdS$_4\times {\rm S}^7$.
We stress that in the general case however the metric (\ref{internalG_intro}) 
will not be part of a solution of the higher-dimensional field equations. 
This is equivalent to the fact that the lower-dimensional supergravities
in general do not have a critical point at the origin of the scalar potential, as explicitly verified
in~\cite{Hull:1988jw} for the ${\rm SO}(p,8-p)$ supergravities.
Nevertheless, in all cases the generalized Scherk-Schwarz 
ansatz continues to describe the consistent truncation of the higher-dimensional supergravity 
to the field content and the dynamics of a
lower-dimensional maximally supersymmetric supergravity.
The construction thus enriches the class of known consistent truncations 
not only by the long-standing AdS$_5\times {\rm S}^5$, but also by various
hyperboloid compactifications giving rise to non-compact and 
non-semisimple gauge groups.

\addtocounter{footnote}{-1}
Let us stress that throughout this paper we impose the strong version of the section 
constraint, which implies that locally the fields (i.e.\ the twist matrix $U$ and scale factor $\rho$) 
depend only on the coordinates of the usual supergravities.\footnotemark\
This is indispensable in order to deduce that the consistent truncations from exceptional
field theory induce a consistent truncation of the original supergravities.
On the other hand, it puts additional constraints on the solutions of 
(\ref{consistent_intro}), which makes the actual construction of such solutions 
a more difficult task. 
Although naively, one might have thought that for a given embedding tensor $\Theta_{M}{}^{\bfa}$ 
a simple exponentiation of $Y^M\Theta_M{}^{{\bfa}} t_{\bfa}$ provides a candidate for
a proper twist matrix, the failure of Jacobi identities of the `structure constants' 
$\Theta_M{}^{{\bfa}} (t_{\bfa})_N{}^K$,
and the non-trivial projection in (\ref{consistent_intro}) put a first obstacle to the naive
extrapolation of the Lie algebra structures underlying the standard Scherk-Schwarz ansatz.
On top of this, an object like $Y^M\Theta_M{}^{{\bfa}} t_{\bfa}$ in general violates
the section constraint, since $\Theta_M{}^{\bfa}$ in general
will have a rank higher than is permitted by the six/seven coordinates among the $Y^M$ 
that are consistent with the section constraint. From this point of view,
the standard sphere compactifications take a highly `non-geometric' form. 
While we do not expect to encode in this construction genuinely 
non-geometric compactifications (unless global issues of the type addressed in 
double field theory in \cite{Hohm:2013bwa} become important), we expect that 
a proper understanding of highly non-trivial compatifications like for spheres and hyperboloides
will be a first step in developing a proper conceptual framework for non-geometric compatifications, 
which so far are out of reach. 
It should be evident that the advantage even of the strongly constrained exceptional field theory formulations is a 
dramatic technical simplification of, for instance, the issues related to consistency proofs, 
allowing to resolve old and new open questions. 
In fact, with the full EFTs at hand we can potentially 
provide a long list of examples of consistent truncations that were previously considered unlikely, 
such as hyperboloides, warped spheres, compactifications with massive multiplets, etc.
Of course, eventually one would like to also include in a consistent framework
truly non-geometric compactifications, pointing to a possible relaxation of 
the strong form of the section constraint.

The rest of this paper is organized as follows. 
In section~\ref{sec:eftrev}, we give a brief review of the ${\rm E}_{7(7)}$ EFT.
Although in this paper most detailed technical discussions will be presented for the ${\rm E}_{7(7)}$ EFT
the analogous constructions go through for all other EFTs. 
In section~\ref{sec:reduction} we describe the generalized Scherk-Schwarz ansatz for the
full field content of the theory. We show that it defines a consistent truncation of the EFT 
which reduces to the complete set of field equations of lower-dimensional gauged supergravity with
embedding tensor $\Theta_M{}^\bfa$, $\vartheta_M$, even in presence of a trombone gauging $\vartheta_M\neq0$\,.
For $\vartheta_M=0$, the reduction is also consistent on the level of the action. 
In section~\ref{sec:sphere}, we construct twist matrices $U$ as explicit solutions of the consistency conditions
(\ref{consistent_intro}). We define a family of ${\rm SL}(n)$ twist matrices  and show that upon proper embedding
into the exceptional groups they solve equations (\ref{consistent_intro}). The lower-dimensional theories
have gauge groups ${\rm SO}(p,q)$ and ${\rm CSO}(p,q,r)$, respectively. The construction provides
the consistent reduction ansaetze for compactifications around spheres $S^{n-1}$ and (warped) hyperboloides
$H^{p,q}$\,. Discussion and outlook are given in section~\ref{sec:conclusions}.


\section{E$_{7(7)}$ exceptional field theory}
\label{sec:eftrev}


We start by giving a brief review of the E$_{7(7)}$-covariant exceptional field theory,
constructed in Refs.~\cite{Hohm:2013pua,Hohm:2013uia,Godazgar:2014nqa} (to which we refer for details)\,. 
All fields in this theory depend on the four external variables $x^\mu$, $\mu=0, 1, \dots, 3$, and the 
56 internal variables $Y^M$, $M=1, \dots, 56$, transforming in
the fundamental representation of ${\rm E}_{7(7)}$,
however the latter dependence is strongly restricted by the section condition~\cite{Berman:2010is,Coimbra:2011ky,Berman:2012vc}
\be
 \begin{split}
  (t_\bfa)^{MN}\,\partial_M \otimes \partial_N \ &\equiv \ 0\;, \qquad  
  \Omega^{MN}\,\partial_M\,\otimes \partial_N  \ \equiv \ 0 \,, 
 \end{split}
 \label{sectioncondition}
 \ee  
 where the notation $\otimes$ should indicate that both derivative operators may act on different fields.
 Here, $\Omega^{MN}$ is the symplectic invariant matrix which we use for lowering and raising
of fundamental indices according to $X^M=\Omega^{MN}X_N$, $X_N=X^M\Omega_{MN}$\,.
The tensor $(t_\bfa)_M{}^{N}$ is the representation matrix of ${\rm E}_{7(7)}$ in the fundamental representation,
$\bfa=1, \dots, 133$.
These constraints admit (at least) two inequivalent solutions, in which the fields
depend on a subset of seven or six of the internal variables.
The resulting theories are the full $D=11$ supergravity and the type IIB theory, respectively.

\subsection{Bosonic field equations}

The bosonic field content of the E$_{7(7)}$-covariant exceptional field theory is given by 
\bea
\left\{e_\mu{}^{\alpha}\,,\; {\cal M}_{MN}\,,\; \cA_\mu{}^M\,,\; \cB_{\mu\nu\,\bfa}\,,\; 
\cB_{\mu\nu\,M} \right\}
\;.
\label{fieldcontent}
\eea
The field $e_\mu{}^{\alpha}$ is the vierbein, from which the external (four-dimensional) 
metric is obtained as $g_{\mu\nu}=e_\mu{}^{\alpha} e_{\nu \alpha}$.
The scalar fields are described by the symmetric matrix ${\cal M}_{MN}$ constructed as
${\cal M}_{MN}=({\cal V}{\cal V}^T)_{MN}$ from an ${\rm E}_{7(7)}$ valued 56-bein,
parametrizing the coset space ${\rm E}_{7(7)}/{\rm SU}(8)$.
Vectors $\cA_\mu{}^M$ and 2-forms $\cB_{\mu\nu\,\bfa}$ transform in the fundamental
and adjoint representation of ${\rm E}_{7(7)}$, respectively. 
The 2-forms $\cB_{\mu\nu\,N}$ in the fundamental representation describe a covariantly constrained tensor field,
i.e.\ satisfy algebraic equations analogous to (\ref{sectioncondition})
\be
 \begin{split}
  (t_\bfa)^{MN}\,\cB_{M } \otimes \cB_{N} \ &= \ 0\;, \qquad 
  (t_\bfa)^{MN}\,\cB_M\otimes \partial_N \ = \ 0  \,, \\
  &\Omega^{MN}\,\cB_M \otimes \cB_N\ = \ 0 \,, \qquad \Omega^{MN}\,\cB_M\otimes \partial_N\ = \ 0 \,.
 \end{split}
 \label{sectionconditionB}
 \ee  
Their presence is necessary for consistency of the hierarchy of non-abelian 
gauge transformations and can be inferred directly from the properties of
the Jacobiator of generalized diffeomorphisms~\cite{Hohm:2013uia}.
In turn, after solving the section constraint these fields ensure the correct and duality covariant
description of those degrees of freedom that are on-shell dual to the higher-dimensional gravitational degrees of freedom. 

The bosonic exceptional field theory is invariant under generalized diffeomorphisms in the internal
coordinates, acting via~\cite{Coimbra:2011ky}
\bea\label{genLie}
\mathbb{L}_{\Lambda} 
& \equiv & \Lambda^K \partial_K  + 12\, \partial_K \Lambda^L\,(t^\bfa)_L{}^K\,t_\bfa
+\lambda\,\partial_P \Lambda^P
\;,
\eea
on arbitrary ${\rm E}_{7(7)}$ tensors of weight $\lambda$.
The weights of the various bosonic fields of the theory are given by
\bea
\begin{array}{c||c|c|c|c|c}
 & e_{\mu}{}^{\alpha} & {\cal M}_{MN}  & \cA_\mu{}^M &
\cB_{\mu\nu\,\bfa} &
\cB_{\mu\nu\,M}
\\ \hline
\lambda \;:\; & \frac12 & 0 & \frac12 & 1 & \frac12  \\
\end{array}\quad
\;.
\label{weights}
\eea
The generalized diffeomorphisms  give rise to the
definition of covariant derivatives 
\bea
{\cal D}_\mu=\partial_\mu - \mathbb{L}_{\cA_\mu}
\;,
\label{covD}
\eea
covariantizing the theory under $x$-dependent transformations~(\ref{genLie}).
Their commutator closes into the non-abelian field strengths ${\cal F}_{\mu\nu}{}^M$ defined by
\bea
{\cal F}_{\mu\nu}{}^M &\equiv&  2 \partial_{[\mu} \cA_{\nu]}{}^M 
-2\,\cA_{[\mu}{}^N \partial_N \cA_{\nu]}{}^M 
-\frac1{2}\left(24\, (t_\bfa)^{MN} (t^\bfa)_{KL}
-\Omega^{MN}\Omega_{KL}\right)\,\cA_{[\mu}{}^K\,\partial_N \cA_{\nu]}{}^L
\nonumber\\
&&{}
- 12 \,  (t^\bfa)^{MN} \,\partial_N \cB_{\mu\nu\,\bfa}
-\frac12\,\Omega^{MN}\,\cB_{\mu\nu\,N}
\;.
\label{YM}
\eea
The two-forms $\cB_{\mu\nu\,\bfa}$, $\cB_{\mu\nu\,N}$ drop out from the commutator $[{\cal D}_{\mu}, {\cal D}_{\nu}]$
but are required for ${\cal F}_{\mu\nu}{}^M$ to transform covariantly under generalized diffeomorphisms.

The equations of motion of the bosonic theory
are most compactly described by a Lagrangian
\bea
\label{finalaction}
 {\cal L}_{\rm \eft}&=&  e\,  \widehat{R}
 +\frac{1}{48}\,e\,g^{\mu\nu}\,{\cal D}_{\mu}{\cal M}^{MN}\,{\cal D}_{\nu}{\cal M}_{MN}
  -\frac{1}{8}\,e\,{\cal M}_{MN}\,{\cal F}^{\mu\nu M}{\cal F}_{\mu\nu}{}^N
 \nonumber\\&&{}
+{\cal L}_{\rm top}-e\,V({\cal M}_{MN},g_{\mu\nu}) \,.
\eea
Let us review the different terms. The modified Einstein Hilbert term carries the Ricci scalar $\widehat{R}$ obtained
from contracting the modified Riemann tensor
  \be
  \widehat{R}_{\mu\nu}{}^{\alpha \beta} \ \equiv \  R_{\mu\nu}{}^{\alpha \beta}[\omega]+{\cal F}_{\mu\nu}{}^{M}
  e^{\alpha}{}^{\rho}\partial_M e_{\rho}{}^{\beta}\,,
  \label{improvedRE7}
 \ee
with the spin connection $\omega_\mu{}^{\alpha \beta}$ obtained from the covariantized vanishing torsion condition
${\cal D}_{[\mu} e_{\nu]}{}^{\alpha} \equiv0$\,.
Scalar and vector kinetic terms are defined in terms of the covariant derivatives (\ref{covD}) and 
field strengths (\ref{YM}).
The Lagrangian (\ref{finalaction}) is to be understood as a ``pseudo-Lagrangian"
in the sense of a democratic action~\cite{Bergshoeff:2001pv},
with the vector fields further subject to the first order duality equations 
\bea
{\cal F}_{\mu\nu}{}^M &=&  
\frac12\,i\,e\varepsilon_{\mu\nu\rho\sigma}\,\Omega^{MN} {\cal M}_{NK}\,{\cal F}^{\rho\sigma}{}^K
\;,
\label{dualityM}
\eea
to be imposed after varying the second-order Lagrangian.
The topological term in (\ref{finalaction}) is most compactly given as the boundary 
contribution of a five-dimensional bulk integral
 \bea
 \label{Ltop}
\int_{\partial\Sigma_5} d^4x\,  \int d^{56}Y\, {\cal L}_{\rm top} 
&=& \frac{i}{24}\,\int_{\Sigma_5} d^5x \,\int d^{56}Y\,
\varepsilon^{\mu\nu\rho\sigma\tau}\,{\cal F}_{\mu\nu}{}^M\,
{\cal D}_{\rho} {\cal F}_{\sigma\tau}{}_M
\;.
 \eea
Finally, the last term in (\ref{finalaction}) is given by
\bea
  V({\cal M}_{MN},g_{\mu\nu})  &=& 
  -\frac{1}{48}{\cal M}^{MN}\partial_M{\cal M}^{KL}\,\partial_N{\cal M}_{KL}+\frac{1}{2} {\cal M}^{MN}\partial_M{\cal M}^{KL}\partial_L{\cal M}_{NK}\label{fullpotential}\\
  &&{}-\frac{1}{2}g^{-1}\partial_Mg\,\partial_N{\cal M}^{MN}-\frac{1}{4}  {\cal M}^{MN}g^{-1}\partial_Mg\,g^{-1}\partial_Ng
  -\frac{1}{4}{\cal M}^{MN}\partial_Mg^{\mu\nu}\partial_N g_{\mu\nu}\;,
\nonumber 
\eea 
in terms of the internal and external metric. 

The full bosonic theory is invariant under vector and tensor gauge symmetries
with parameters $\Lambda^M$, $\Xi_{\mu \, \bfa}$, $\Xi_{\mu \, M}$ (the latter constrained according to (\ref{sectionconditionB})), as well as under generalized diffeomorphisms in the external coordinates.
Together, these symmetries uniquely fix all field equations.

\subsection{SU$(8)\times {\rm E}_{7(7)}$ Geometry and Fermions}\label{E7GEOMetry}
In this subsection we review some aspects of the SU$(8)\times {\rm E}_{7(7)}$ covariant 
geometry, which will be instrumental below, and introduce the fermions of the supersymmetric theory. 
The geometry, which 
closely follows that of double field theory \cite{Siegel:1993th,Hohm:2010xe}, 
was developed for the fields truncated to the internal sector 
in \cite{Coimbra:2011ky,Coimbra:2012af,Aldazabal:2013mya,Cederwall:2013naa} 
and recently completed in \cite{Godazgar:2014nqa} for the full E$_{7(7)}$ exceptional field theory 
constructed in \cite{Hohm:2013uia}. 

We start by introducing a frame formalism, in which the generalized metric  
is expressed in terms of an E$_{7(7)}$ valued vielbein ${\cal V}$, 
 \be\label{MFRame} 
  {\cal M}_{MN} \ = \ {\cal V}_{M}{}^{\underline{A}}\,{\cal V}_{N}{}^{\underline{B}}\,\delta_{\underline{A}\underline{B}}
  \ \equiv \ {\cal V}_{Mij}{\cal V}_{N}{}^{ij}+{\cal V}_{M}{}^{ij}{\cal V}_{Nij}\;, 
 \ee
with flat SU$(8)$ indices $\underline{A},\underline{B},\ldots =1,\ldots, 56$ that split according to the embedding 
$56=28+28$ as $\underline{A} = (\,{}_{[ij]}\, ,\; {}^{[ij]} )$, $i,j=1,\ldots,8$. Formulated in terms of ${\cal V}$, the 
theory exhibits a local SU$(8)$ `tangent space' symmetry. Consequently, one can 
introduce connections ${\cal Q}_M$ for this symmetry that can be expressed as 
 \be
  {\cal Q}_{M\underline{A}}{}^{\,\underline{B}} \ = \ {\cal Q}_{M ij}{}^{kl} \ = \ {\cal Q}_{M [i}{}^{[k}\delta_{j]}{}^{l]} 
  \label{Qij}
 \ee 
in terms of SU$(8)$ connections ${\cal Q}_{Mi}{}^{j}$ with fundamental indices. We may also introduce a Christoffel-type 
connection $\Gamma_{MN}{}^{K}$ and (internal) spin connections $\omega_M{}^{\alpha\beta}$ that render 
derivatives covariant
under generalized diffeomorphisms and local SO$(1,3)$ transformations, respectively.  
On a generalized vector $V_{Mi}$ transforming 
in the fundamental of E$_{7(7)}$ and SU$(8)$ and as a spinor under SO$(1,3)$ 
(whose spinor index we suppress) the covariant derivative is given by 
 \be
 \begin{split}
  \nabla_MV_{Ni} \ \equiv \ & \,\partial_MV_{Ni}+\frac{1}{4}\omega_{M}{}^{\alpha\beta}\gamma_{\alpha\beta}V_{Ni}
  +\frac{1}{2}{\cal Q}_{M i}{}^{j} V_{Nj}\\
  &-\Gamma_{MN}{}^{K} V_{Ki}-\frac{2}{3}\lambda(V)\Gamma_{KM}{}^{K} V_{Ni}\;, 
 \end{split} 
 \label{nabla}
 \ee 
where $\lambda$ is the density weight of $V$. 
The internal SO$(1,3)$ spin connection is given by 
 \be
  \omega_{M}{}^{\alpha\beta} \ = \ e^{\mu[\alpha}\partial_M e_{\mu}{}^{\beta]}\;. 
 \ee
 Sometimes it is convenient to work with the combination 
 \be
  \widehat{\omega}_{M}{}^{\alpha\beta} \ \equiv \    \omega_{M}{}^{\alpha\beta}
  -\frac{1}{4}{\cal M}_{MN}{\cal F}_{\mu\nu}{}^{N} e^{\mu\alpha} e^{\nu\beta}\;, 
 \ee
which naturally enters the supersymmetry variations to be given momentarily. 

Next, the remaining connections in (\ref{nabla}) can be determined (in part) in terms of the physical fields
by imposing further constraints. While this does not determine all connections uniquely, the 
undetermined connection components drop out of all relevant  
expressions. The first constraint is the generalized torsion constraint, setting to zero a generalized 
torsion tensor ${\cal T}_{MN}{}^{K}$. In order to state this constraint, note that 
the Christoffel connection, in its last two indices, takes values 
in the 133-dimensional Lie algebra of E$_{7(7)}$, 
and therefore these connections live in 
the representation
 \be
  \Gamma_{MN}{}^{K}\,:\qquad {\bf 56}\otimes {\bf 133} \ = \ {\bf 56}\oplus {\bf 912}\oplus {\bf 6480}\;. 
 \ee 
The torsion constraint simply sets the ${\bf 912}$ sub-representation to zero, 
 \be\label{GENTOR}
  {\cal T}_{MN}{}^{K} \ = \ 0\quad\Longleftrightarrow \quad \Gamma_{MN}{}^{K}\Big|_{\bf 912} \ = \ 0\;,  
 \ee  
which may be verified to be a gauge covariant condition. 
Next, demanding that the covariant derivative is compatible with the vierbein density 
$e=\det e_{\mu}{}^{\alpha}$ fixes 
 \be\label{TRACEconstr}
  \nabla_M e\ = \ 0\qquad \Longleftrightarrow \qquad
  \Gamma_{KM}{}^{K} \ = \ \frac{3}{4}e^{-1}\partial_Me\;, 
 \ee 
allowing for integration by parts with covariant derivatives. 
This determines the ${\bf 56}$ part of $\Gamma$. 
It implies, for instance, that the covariant derivative of 
an E$_{7(7)}$ singlet with density weight $\lambda$ in the fundamental 
of SU$(8)$ can be written as  
 \be
  \nabla_M V_i \ = \ e^{\frac{\lambda}{2}}\, \partial_M \big(e^{-\frac{\lambda}{2}}\,V_i\big)
  +\frac{1}{2}{\cal Q}_{Mi}{}^{j} V_j\;. 
 \ee  
The final constraint is the `vielbein postulate' stating that the frame field is covariantly 
constant w.r.t.~the combined action of the Christoffel and SU$(8)$ 
connection, 
 \be\label{vielbeinpost}
  \nabla_M{\cal V}{}_{N}{}^{\underline{A}} \ \equiv \ \partial_M{\cal V}_N{}^{\underline{A}}-
  {\cal Q}_{M\underline{B}}{}^{\underline{A}}{\cal V}_{N}{}^{\underline{B}}
  -\Gamma_{MN}{}^{K}{\cal V}_{K}{}^{\underline{A}} \ = \ 0\;, 
 \ee 
or, in terms of fundamental SU$(8)$ indices, 
 \be
   \nabla_M{\cal V}{}_{N}{}^{ij} \ \equiv \ \partial_M{\cal V}_N{}^{ij}+{\cal Q}_{Mk}{}^{[i}{\cal V}_{N}{}^{j]k}
  -\Gamma_{MN}{}^{K}{\cal V}_{K}{}^{ij} \ = \ 0\;, 
 \ee
and similarly for its complex conjugate with lower indices.  
 
Let us now give explicit expressions for the determined parts of the spin connections 
that we will use below. We first note that the vielbein postulate relates the Christoffel connection to the SU$(8)$ 
connection as 
 \be
  \Gamma_{\underline{A}\underline{B}}{}^{\,\underline{C}} 
  \ \equiv \ ({\cal V}^{-1})_{\underline{A}}{}^{M}({\cal V}^{-1})_{\underline{B}}{}^{N}
  \Gamma_{MN}{}^{K}{\cal V}_{K}{}^{\underline{C}}
  \ = \ ({\cal V}^{-1})_{\underline{A}}{}^{M}({\cal V}^{-1})_{\underline{B}}{}^{N}\partial_M{\cal V}_{N}{}^{\underline{C}}
  -{\cal Q}_{\underline{A}\underline{B}}{}^{\,\underline{C}}\;, 
 \ee 
where the indices on $\Gamma$ are `flattened' by means of the frame field ${\cal V}$. 
Projecting both sides of this equation onto the ${\bf 912}$, the 
generalized torsion constraint (\ref{GENTOR}) implies that
  \be\label{detConn1}
  \Big[{\cal Q}_{\underline{A}\underline{B}}{}^{\,\underline{C}} \Big]_{\bf 912} \ = \ 
  \Big[({\cal V}^{-1})_{\underline{A}}{}^{M}({\cal V}^{-1})_{\underline{B}}{}^{N}\partial_M{\cal V}_{N}{}^{\underline{C}}
    \Big]_{\bf 912}\;. 
 \ee
Thus, while the ${\bf 912}$ projection of the Christoffel connection is set to zero by the 
torsion constraint, the  ${\bf 912}$ projection of the SU$(8)$ connection is precisely the part 
determined by the torsion constraint. 
Similarly, one obtains an expression for the trace part of ${\cal Q}$. 
 Taking the trace of  (\ref{vielbeinpost}) yields
 \be
  \Gamma_{MN}{}^{M} \ = \ {\cal V}_{N}{}^{\underline{A}}\big(-\partial_M{\cal V}_{\underline{A}}{}^{M}
  +{\cal Q}_{\underline{B}\underline{A}}{}^{\underline{B}}\big)\;. 
 \ee 
Inserting (\ref{TRACEconstr}) this implies 
 \be\label{detConn2}
 {\cal Q}_{\underline{B}\underline{A}}{}^{\underline{B}} 
  \ = \ \partial_M{\cal V}_{\underline{A}}{}^{M}+\frac{3}{4}e^{-1}{\cal V}_{\underline{A}}{}^{M}\partial_M e\;. 
 \ee 
Eqs.~(\ref{detConn1}) and (\ref{detConn2}) give the full part of the SU$(8)$ connection 
that is determined by the above constraints. 

We close the discussion of the geometry by giving a definition of the 
generalized scalar curvature ${\cal R}$ 
that enters the potential. It can be defined 
through a particular combination of second-order terms in covariant derivatives 
acting on an SO$(1,3)$ spinor in the fundamental of SU$(8)$ with density weight $\frac{1}{4}$
(that below will be identified with the supersymmetry parameter),    
 \be\label{covPOT}
   6\, \nabla_{ik}\nabla^{kj}\epsilon_j
  +2\, \nabla^{kj}\nabla_{ik}\epsilon_j
  + \nabla^{kl}\nabla_{kl}\epsilon_i 
  \ = \ 
   \frac{1}{16} {\cal R}\epsilon_i 
   -\frac{1}{4}\gamma^{\mu\nu}g^{\rho\sigma}\nabla_{ik}g_{\mu\rho}\,\nabla^{kj}g_{\nu\sigma}\epsilon_j\;. 
  \ee 
Here covariant derivative with flattened index are defined as
 \be\label{flattened}
  \nabla^{ij} \ \equiv \ ({\cal V}^{-1})^{ij\,M} \, \nabla_{M}\;. 
 \ee 
In (\ref{covPOT}) all undetermined connections drop out. Thus,  
the ${\cal R}$ defined by this equation is a gauge scalar, 
which enters the potential (\ref{fullpotential}) as 
 \be\label{scalarPot}
  V \ = \ -{\cal R} \, - \, \frac{1}{4}{\cal M}^{MN}\nabla_M g^{\mu\nu}\,\nabla_N g_{\mu\nu}\;.
 \ee 

\medskip

Let us finally turn to the fermions, which are scalar densities under 
internal generalized diffeomorphisms but transform in non-trivial 
representations of the tangent space symmetries SO$(1,3)$ and SU$(8)$. 
The gravitino reads $\psi_{\mu}{}^{i}$, and it has the same weight
as the supersymmetry parameter $\epsilon$:
 \be
  \lambda(\psi_{\mu}{}^{i}) \ = \ \lambda(\epsilon) \ = \ \frac{1}{4}\;. 
 \ee
The 56 `spin-$\tfrac{1}{2}$' fermions are given by $\chi^{ijk}$, 
totally antisymmetric in their SU$(8)$ indices,  with density weight 
 \be
  \lambda(\chi^{ijk}) \ = \ -\frac{1}{4}\;. 
 \ee 
The supersymmetry variations take the manifestly covariant form written 
in terms of the above covariant derivatives
 \be\label{fullSUSY}
 \begin{split}
  \delta_{\epsilon}\psi_{\mu}{}^{i} \ &= \ 2\,{\cal D}_{\mu}\epsilon^i
  -4\,\widehat{\nabla}^{ij}\big(\gamma_{\mu}\epsilon_j\big)\;, \\
  \delta_{\epsilon}\chi^{ijk} \ &= \ -2\sqrt{2}\,{\cal P}_{\mu}{}^{ijkl}
  \gamma^{\mu}\epsilon_l-12\sqrt{2}\,\widehat{\nabla}^{[ij}\,\epsilon^{k]} \;. 
 \end{split}
 \ee 
Here ${\cal P}_{\mu}$ is the non-compact part of the covariantized Lie algebra valued 
current ${\cal V}^{-1}{\cal D}_{\mu}{\cal V}$, whose precise form is not 
important for our purposes in this paper (see \cite{Godazgar:2014nqa} for 
the definition).  
The projections or contractions of covariant derivatives in these supersymmetry 
variations  are again 
such that the undetermined connections drop out.

\subsection{Solutions of the section condition}

Even though the fields of (\ref{finalaction}) formally depend on $4+56$ coordinates, their dependence
on the internal coordinates is severely restricted by the section constraints~(\ref{sectioncondition}).
These constraints admit (at least) two inequivalent solutions, in which the fields depend on a subset of 
seven or six of the internal variables, respectively, according to the decompositions
of the fundamental representation of ${\rm E}_{7(7)}$ with respect to 
the maximal subgroups ${\rm GL}(7)$ and ${\rm GL}(6)\times {\rm SL}(2)$, 
respectively
\bea
 {\bf 56} &\longrightarrow&
\boxed{{7}_{+3}}+ { 21}'_{+1}+{21}_{-1}+ {7}'_{-3}\;,
\nonumber\\
{\bf 56} &\longrightarrow&
\boxed{({6},1)_{+2}}+(6',2)_{+1}+(20,1)_0 +(6,2)_{-1}+ (6',1)_{-2}
\;.
\label{decIIAB}
\eea
Upon imposing this restricted coordinate dependence on all fields, the 
Lagrangian (\ref{finalaction}) (upon proper dualizations and field redefinitions)
exactly reproduces the full field equations of
$D=11$ supergravity and the type IIB theory, respectively.

The various fields of higher-dimensional supergravity are recovered by splitting the
fields (\ref{fieldcontent}) according to the decompositions (\ref{decIIAB}).
Their higher-dimensional origin is then most conveniently identified via the ${\rm GL}(1)$ grading
that is captured by the subscripts in (\ref{decIIAB}).
E.g.\ the scalar matrix is parametrized as (\ref{MFRame}) in terms of the group-valued vielbein ${\cal V}$, 
parametrized in triangular gauge associated with the ${\rm GL}(1)$ grading \cite{Cremmer:1997ct} 
according to
\bea
{\cal V}_{\rm IIA} &\equiv& {\rm exp} \left[\phi\, t^{\rm IIA}_{(0)}\right]\,{\cal V}_7\;
{\rm exp}\left[c_{kmn}\,t_{(+2)}^{kmn}\right]\,{\rm exp}\left[\epsilon^{klmnpqr}  c_{klmnpq}\, t_{(+4)\,r}\right]
\;,
\label{V56}\\
{\cal V}_{\rm IIB} &\equiv& {\rm exp} \left[\phi\, t^{\rm IIB}_{(0)}\right]\,{\cal V}_6\,{\cal V}_2\;
{\rm exp}\left[c_{mn\,a}\,t_{(+1)}^{mn\,a}\right]\,
{\rm exp}\left[\epsilon^{klmnpq}\,c_{klmn}\,t_{(+2)}{\,}_{pq}\right]
{\rm exp}\left[c_{a}\,t_{(+3)}^{a}\right]
\;,
\nonumber
\eea
see \cite{Hohm:2013uia} for details. 
The dictionary further 
requires redefinitions of all the form fields originating from the higher-dimensional $p$-forms in the usual 
Kaluza-Klein manner, i.e., flattening the world indices with the elfbein and then `un-flattening' with the vierbein 
$e_{\mu}{}^{a}$, as well as subsequent further non-linear field redefinitions and appropriate dualization of
some field components, c.f.~\cite{Hohm:2013vpa,Hohm:2013uia}.

\section{Reduction ansatz}
\label{sec:reduction}

In this section we first review the generalized Scherk-Schwarz ansatz of
\cite{Kaloper:1999yr,Aldazabal:2011nj,Geissbuhler:2011mx,Grana:2012rr,Berman:2012uy,Aldazabal:2013mya}.
We then extend it to the full field content of the exceptional field theory (\ref{finalaction}) and
find in particular, that it requires a non-trivial ansatz for the constrained compensator gauge field $B_{\mu\nu\,M}$
of (\ref{sectionconditionB}). Together, this defines a consistent truncation of the field equations derived 
from (\ref{finalaction}).

\subsection{Generalized Scherk-Schwarz ansatz}

The reduction ansatz for the matrix of scalar fields is given by the matrix product
\bea
{\cal M}_{MN}(x,Y) &=& U_{M}{}^{\underline{K}}(Y)\,U_{N}{}^{\underline{L}}(Y)\,{M}_{\underline{K}\underline{L}}(x)\;,
\label{genSS}
\eea
with an ${\rm E}_{7(7)}$-valued twist matrix $U_M{}^{\underline{N}}$ satisfying the first order differential 
equations~\cite{Aldazabal:2013mya}
\bea
\left[(U^{-1})_{\underline{M}}{}^{P} (U^{-1})_{\underline{N}}{}^{Q} \, \partial_P U_Q{}^{\underline{K}}\right]_{(912)} 
&\stackrel{!}{=}&
\frac1{7}\,\rho \,\Theta_{\underline{M}}{}^\bfa\, (t_\bfa)_{\underline{N}}{}^{\underline{K}}
\;,
\nonumber\\
  \partial_N (U^{-1})_{\underline{M}}{}^N  
- 3\,\rho^{-1}\partial_N \rho  \, (U^{-1})_{\underline{M}}{}^N  &\stackrel{!}{=}& 
2\,\rho\,\vartheta_{\underline{M}}
\;,
\label{consistent_twist}
\eea
with a $Y$-dependent factor $\rho(Y)$ and constant tensors $\Theta_{\underline{M}}{}^\bfa$ and 
$\vartheta_{\underline{M}}$.\footnote{
\label{foot:norm}
With respect to the general form (\ref{consistent_intro}) of these equations 
we have introduced a different explicit normalization factor $1/7$ (which could be absorbed into $\Theta_M{}^\bfa$)
to achieve later agreement with the $D=4$ formulas.
Moreover, for matching the conventions of gauged supergravity~\cite{deWit:2007mt,LeDiffon:2011wt} 
we are obliged to perform a rescaling of vector and tensor gauge fields 
$\sqrt{2}\,A_\mu{}^{M}_{[1103.2785]}=A_\mu{}^{M}_{{\rm [here]}}$, $B_{\mu\nu\,\bfa}^{[1103.2785]}=-B_{\mu\nu\,\bfa}^{{\rm [here]}}$,
and accordingly of the embedding tensor $\Theta_M{}^{\bfa}_{[1103.2785]}=\sqrt{2}\, \Theta_M{}^{\bfa}_{{\rm [here]}}$,
$\vartheta_{M}^{[1103.2785]}=\sqrt{2}\, \vartheta_{M}^{{\rm [here]}}$\,.
This is most easily seen by comparing the supersymmetry transformation rules 
from~\cite{deWit:2007mt,LeDiffon:2011wt} to \cite{Godazgar:2014nqa}.
}
The latter can be identified with the embedding tensor of the gauged supergravity to which the theory
reduces after the generalized Scherk-Schwarz ansatz. The notation $[\cdot]_{(912)}$ refers to projection
onto the irreducible ${\bf 912}$ representation of ${\rm E}_{7(7)}$.
The density factor $\rho(Y)$ has no analogue in the standard Scherk-Schwarz reduction \cite{Scherk:1979zr}
but ensures that the consistency equations (\ref{consistent_twist}) transform covariantly under internal generalized 
diffeomorphisms despite the non-trivial weight carried by the internal derivatives.
As can be verified by a direct computation analogous to that proving the covariance of 
the torsion constraint, 
the consistency equations are covariant if $U_{M}{}^{\underline{N}}$ is treated as 
a generalized vector indicated by the index $M$ while the index $\underline{N}$ 
refers to the \textit{global} ($x$ and $Y$ independent) 
E$_{7(7)}$ that is preserved by the Scherk-Schwarz reduction ansatz. 
This is the global E$_{7(7)}$ that is a covariance of the embedding tensor formulation of gauged 
supergravity (and that, for fixed embedding tensor components $\Theta_{\underline{M}}{}^\bfa$ and 
$\vartheta_{\underline{M}}$, is broken to the gauge group). In the following we will not indicate 
the difference between bare and underlined E$_{7(7)}$ indices since the nature of indices can 
always be inferred from their position in $U$ and $U^{-1}$. 

This Scherk-Schwarz reducution ansatz (\ref{genSS})  is made such that the action of generalized diffeomorphisms (\ref{genLie}) with parameter
\bea
\Lambda^M(x,Y) &=& \rho^{-1}(Y)\,(U^{-1})_P{}^M(Y)\,\Lambda^P(x)
\;,
\label{SSL}
\eea
on (\ref{genSS}) is compatible with the reduction ansatz and induces an action
\bea
\delta_\Lambda {M}_{MN}(x) &=& 
2\,\Lambda^L(x)\left(
\Theta_L{}^\bfa
+8\,\vartheta_R\,(t^\bfa)_L{}^R  \right) (t_\bfa)_{(M}{}^P\,{M}_{N)P}(x)
\;,
\label{delM}
\eea
on the $Y$-independent $M_{MN}$ which is precisely the action of gauge transformations in $D=4$ maximal supergravity
with the constant embedding tensor $\Theta_M{}^\bfa$, $\vartheta_M$\,.
The second term in (\ref{delM}) captures the gauging of the trombone scaling symmetry~\cite{LeDiffon:2008sh}.
The consistency conditions (\ref{consistent_twist}) together with the section condition~(\ref{sectioncondition})
for the twist matrix imply that $\Theta_M{}^\bfa$, $\vartheta_M$ automatically satisfy the
quadratic constraints~\cite{deWit:2007mt,LeDiffon:2011wt} that ensure closure of the gauge algebra.

Throughout, we will impose compatibility of the reduction ansatz with the section constraints~(\ref{sectioncondition}).
This translates into further conditions on the twist matrix $U_M{}^N$\,. In the conclusions we comment on the possible
relaxation of these constraints.
In order to describe the consistent truncation of the full exceptional field theory (\ref{finalaction}), the generalized
Scherk-Schwarz ansatz (\ref{genSS}) has to be extended to the remaining fields of the theory
which is straightforward for the vierbein, vector and tensor fields as
 \bea
 e_\mu{}^\alpha(x,Y) &=& \rho^{-1}(Y)\,e_\mu{}^\alpha(x)\;,\nonumber\\
  {\cal A}_{\mu}{}^{M}(x,Y) &=& A_{\mu}{}^{N}(x)(U^{-1})_{N}{}^{M}(Y)\,\rho^{-1}(Y)\;, 
  \nonumber\\
  {\cal B}_{\mu\nu\,\bfa}(x,Y) &=& \,\rho^{-2}(Y) U_\bfa{}^\bfb(Y)\,B_{\mu\nu\,\bfb}(x)
  \;,
  \label{SSeAB}
 \eea
with $U_\bfa{}^\bfb$ denoting the ${\rm E}_{7(7)}$ twist matrix evaluated in the adjoint representation.
All fields thus transform with the twist matrix $U$ acting on their ${\rm E}_{7(7)}$ indices and the power of $\rho^{-2\lambda}$
corresponding to their weight $\lambda$ under generalized diffeomorphisms (\ref{genLie}).
E.g.\ the ansatz for the vierbein (which transforms as a scalar of weight $\lambda=\frac12$ under (\ref{genLie})) ensures that 
under reduction the action of internal generalized diffeomorphisms consistently reduces to
the action of the trombone scaling symmetry in $D=4$ supergravity
\bea
\delta_\Lambda e_\mu{}^\alpha(x,Y) &=& \Lambda^M(x)
\left( \partial_M (\rho^{-1})  + \lambda\,\partial_M \left( \rho^{-1} (U^{-1})_N{}^M\right) \right)e_\mu{}^\alpha(x)\nonumber\\
&=&
\Lambda^M(x)\,\lambda\,\rho^{-1}
\left(  \partial_M (U^{-1})_N{}^M -(1+\lambda^{-1}) \,\rho^{-1} \partial_M\rho \right) e_\mu{}^\alpha(x)
\nonumber\\
&=& \Lambda^M(x)\,\vartheta_M\,e_\mu{}^\alpha(x)
\;.
\label{dele}
\eea
This confirms that a non-vanishing $\vartheta_M$ induces 
a gauging of the rigid scale invariance (trombone) of the 
supergravity equations that scales metric and matter fields 
with proper weights according to (\ref{weights})~\cite{LeDiffon:2008sh}.  
The reduction ansatz for the vector field follows (\ref{SSL}) and ensures that covariant derivatives (\ref{covD})
reduce properly:
\bea
{\cal D}_\mu e_\nu{}^\alpha(x,Y) &=&
\rho^{-1}\left(
\partial_\mu - A_{\mu}{}^{N}  {\vartheta}_{N}
\right) e_\nu{}^\alpha
\;,\\
{\cal D}_\mu  {\cal M}_{MN}(x,Y) &=& 
U_M{}^PU_N{}^Q\,\Big(
\partial_\mu {M}_{PQ} - 2 
A_\mu^L\left(
\Theta_L{}^\bfa\,
+8\,\vartheta_R\,(t^\bfa)_L{}^R\right) (t_\bfa)_{(M}{}^P {M}_{N)P} \Big)
\;.
\nonumber
\eea
On the other hand, a consistent reduction of its non-abelian field strength~(\ref{YM})
requires a
non-trivial ansatz for the constrained compensator field ${\cal B}_{\mu\nu\,M}$
according to 
\bea
{\cal B}_{\mu\nu\,M} ~=~-2\, \rho^{-2}\,(U^{-1})_S{}^P \partial_M U_P{}^R (t^{\bfa}){}_R{}^S\, B_{\mu\nu\,\bfa} 
\;,
\label{SSB}
\eea
which is manifestly compatible with the constraints (\ref{sectionconditionB}) this field has to satisfy.
With this ansatz, it follows that the field strength~(\ref{YM}) takes the form
\bea
{\cal F}_{\mu\nu}{}^M(x,Y) &=&
\rho^{-1}\,(U^{-1})_N{}^M\,\Big\{
2\,\partial_{[\mu} A_{\nu]}{}^N 
+  \Theta_{K}{}^\bfa \,(t_\bfa)_{L}{}^N \,   A_{[\mu}{}^K A_{\nu]}{}^L 
\nonumber\\
&&{}
\qquad\qquad\qquad\quad
-\frac13\, \left( \Omega^{NP}\Omega_{KL}  +4\, \delta_{KL}^{NP}
\right)
\,\vartheta_P
\,A_{[\mu}{}^K\, A_{\nu]}{}^L
\nonumber\\
&&{}
\qquad\qquad\qquad\quad
+
\left(\Theta^N{}^\bfa  -16 \,    (t^\bfa)^{NK} \vartheta_K\right) B_{\mu\nu\,\bfa}
\Big\}
\nonumber\\
&\equiv&
\rho^{-1}\,(U^{-1})_N{}^M\, {\cal F}_{\mu\nu}{}^N(x)
\;,
\eea
with the $Y$-independent 
${\cal F}_{\mu\nu}{}^N(x)$ precisely reproducing the field strength of general gauged 
supergravity with embedding tensor $\Theta_M{}^\bfa$ and $\vartheta_M$~\cite{deWit:2007mt,LeDiffon:2011wt}
(upon the rescaling of footnote~\ref{foot:norm}).

Let us finally note that
the consistency equations (\ref{consistent_twist}) for the twist matrix $U$ 
are easily generalized to other dimensions as
\bea
\left[(U^{-1})_M{}^P (U^{-1})_N{}^L \, \partial_P U_L{}^K\right]_{(\mathbb{P})} &\stackrel{!}{=}&
\rho \,\Theta_M{}^\bfa\, (t_\bfa)_N{}^K
\;,
\nonumber\\
  \partial_N (U^{-1})_M{}^N  
- (D-1)\,\rho^{-1}\partial_N \rho  \, (U^{-1})_M{}^N  &\stackrel{!}{=}& 
\rho\,(D-2)\,\vartheta_M
\;,
\label{consistent_general}
\eea
up to possible normalization factors that can be absorbed into $\Theta_M{}^\bfa$ and $\vartheta_M$\,.
Here, $D$ denotes the number of external space-time dimensions,
and $[\cdot]_{({\mathbb{P}})}$ denotes the projection onto the representation of the 
corresponding embedding tensor in the $D$-dimensional gauged supergravity, c.f.~\cite{deWit:2002vt}.
The coefficients in the second equation can be extracted from (\ref{dele}) taking into
account that the generic weight of the vielbein is $\lambda=\frac1{D-2}$.
Yet another way of presenting the consistency equations (\ref{consistent_general}) is
\bea
\left[{\cal X}_{M}{}^\bfa \right]_{(\mathbb{P})} &\stackrel{!}{=}&
\rho \,\Theta_M{}^\bfa
\;,
\label{consistent_yet}\\
  {\cal X}_{KM}{}^K 
  &\stackrel{!}{=}& 
(1-D)\,\rho^{-1}\partial_N \rho  \, (U^{-1})_M{}^N+\rho\,(2-D)\,\vartheta_M
\;,\nonumber
\eea
with 
\bea
{\cal X}_{MN}{}^K &\equiv& (U^{-1})_M{}^P (U^{-1})_N{}^L \, \partial_P U_L{}^K ~\equiv~ 
{\cal X}_{M}{}^\bfa\,(t_\bfa)_{N}{}^K
\;.
\label{XXX}
\eea

The reduction ansatz generalises accordingly: all fields whose gauge parameters transform as tensors
under internal generalized diffeomorphisms reduce analogous to (\ref{SSeAB}) with the action of the twist matrix $U$ in
the corresponding group representation and the factor $\rho^{-(D-2)\,\lambda}$ taking care of the weight
under generalized diffeomorphisms.
The additional constrained compensator fields reduce with an ansatz analogous to (\ref{SSB}).
These fields appear among the $(D-2)$-forms of the theory, i.e.\ among the two-forms in $D=4$, ${\rm E}_{7(7)}$ EFT,
c.f.~(\ref{sectionconditionB}), and
among the vector fields in $D=3$, ${\rm E}_{8(8)}$ \eft. For $D>4$ these fields do not enter the Lagrangian, although they
can be defined on-shell through their duality equations.

\subsection{Consistent truncation, fermion shifts and scalar potential}
In the previous subsection we have shown that (external) covariant derivatives 
and field strengths reduce `covariantly' under the generalized Scherk-Schwarz reduction, by which we mean that 
all E$_{7(7)}$ indices are simply `rotated' by the twist matrices, up to an overall scaling by $\rho(Y)$
that is determined by the density weight of the object considered. 
For this to happen it was crucial to include the covariantly constrained compensator 
2-form field with its non-standard Scherk-Schwarz ansatz (\ref{SSB}). 
From these results it follows that in the action, equations of motion and symmetry variations all
$Y$-dependence consistently factors out for all contributions built from \textit{external} 
covariant derivatives ${\cal D}_{\mu}$. The remaining equations reproduce those
of $D$-dimensional gauged supergravity.
It remains to establish the same for contributions defined in terms of the 
\textit{internal} covariant derivatives $\nabla_M$ reviewed in sec.~\ref{E7GEOMetry}. 
These include the scalar potential (\ref{fullpotential}) and the corresponding 
terms in the supersymmetry variations of the fermions (\ref{fullSUSY}).
In the following we verify that all these terms
reduce `covariantly' as well and show that the reduction precisely reproduces
the known scalar potential and fermion shifts in the supersymmetry variations 
of gauged supergravity. 

Before proceeding, let us discuss in a little more detail the 
consistency of the Scherk-Schwarz reduction 
at the level of the action. In fact, the previous argument only shows that the reduction
is consistent at the level of the equations of motion.  
Consistency at the level of the action requires in addition that the embedding tensor 
$\vartheta_M$ in (\ref{consistent_general}) inducing the trombone gauging vanishes. 
This is in precise agreement with the fact that for lower-dimensional trombone gauged 
supergravity there is no action principle. 

In order to illustrate this point, let us consider the covariant divergence of a generic vector 
$W^M$ of weight $\lambda$ which using (\ref{TRACEconstr}) takes the form 
 \be
  \nabla_M W^M \ = \ \partial_M W^M+\tfrac{3}{4}\big(1-\tfrac{2}{3}\lambda\big)e^{-1}\partial_Me \,W^M\;. 
 \ee 
Next we compute the Scherk-Schwarz reduction of this divergence.  
Here and in the following it will be convenient to use the bracket notation 
$\langle\;\rangle$ to indicate that an object is subjected to the Scherk-Schwarz reduction. 
Recalling $\langle e\rangle=\rho^{-4}e$, we compute 
 \be\label{covDivSS}
 \begin{split}
   \big\langle\, \nabla_MW^M \,\big\rangle  \ &= \ \partial_M\big(\rho^{-2\lambda}(U^{-1})_N{}^{M}\big)W^N
   +\tfrac{3}{4}\big(1-\tfrac{2}{3}\lambda\big)\rho^{4}\partial_M\rho^{-4}\,\rho^{-2\lambda}(U^{-1})_N{}^{M} W^N   \\
   \ &= \ \rho^{-2\lambda}\Big[\, \partial_M(U^{-1})_N{}^{M}-3\rho^{-1}\partial_M\rho\,(U^{-1})_N{}^{M}\,\Big] W^N\\
   \ &= \ 2\rho^{1-2\lambda}\, \vartheta_M W^M\;, 
  \end{split}
 \ee  
using (\ref{consistent_twist}) to identify the trombone embedding tensor $\vartheta_M$. 
On the other hand, calculating the equations of motion from the Lagrangian requires 
partial integration with general currents $J^M$ of weight $-1/2$ 
\bea
\int \partial_M (e\,J^M)&=& \int e\,\nabla_M J^M\;,
\label{boundary}
\eea
whose boundary contribution is neglected, in obvious contradiction with (\ref{covDivSS})
unless $\vartheta_M=0$\,.
For non-vanishing trombone parameter $\vartheta_M$, 
the Scherk-Schwarz ansatz thus continues to define a consistent truncation 
on the level of the equations of motion, however the lower-dimensional equations
of motion do not allow for an action principle due to the ambiguities arising from (\ref{boundary}).
The resulting structure corresponds to the trombone gaugings of~\cite{LeDiffon:2008sh,LeDiffon:2011wt}.
This is the analogue of the unimodularity condition $f_{MN}{}^M=0$ to be imposed in standard
Scherk-Schwarz reductions \cite{Scherk:1979zr} for invariance of the measure, c.f.~(\ref{consistent_yet}).
Also in that case a non-vanishing $f_{MN}{}^M\equiv \vartheta_N$ does in fact not spoil the consistency
of the reduction ansatz but just the existence of a lower-dimensional action.

Let us now return to the more general discussion of the Scherk-Schwarz 
reduction of the internal covariant derivatives $\nabla_M$. 
We begin by applying the generalized Scherk-Schwarz compactification 
to the SU$(8)$ connections. 
Applied to the projection (\ref{detConn1}) of the SU$(8)$ connection we obtain 
 \be
 \begin{split}
  \big\langle\, \big[\,{\cal Q}_{\underline{A}\underline{B}}{}^{\,\underline{C}}\,\big]_{\rm 912}\,\big\rangle 
  \ &= \ ({\cal V}^{-1})_{\underline{A}}{}^{{M}}({\cal V}^{-1})_{\underline{B}}{}^{{N}}
  {\cal V}_{{K}}{}^{\underline{C}}\,\Big[(U^{-1})_{{M}}{}^{M} (U^{-1})_{{N}}{}^{N}
  \partial_MU_{N}{}^{{K}}\Big]_{\rm 912}\\
  \ &= \ \frac{1}{7}\,\rho\, ({\cal V}^{-1})_{\underline{A}}{}^{{M}}({\cal V}^{-1})_{\underline{B}}{}^{{N}} 
  {\cal V}_{{K}}{}^{\underline{C}}\,
  \Theta_{{M}}{}^{\,\bfa}\,t_{\bfa\, {N}}{}^{\,{K}}\;, 
 \end{split}
 \ee 
upon using (\ref{consistent_twist}). This expression features the flattened 
version of the embedding tensor, also known as the `T-tensor' \cite{deWit:1982ig} in gauged supergravity. 
Similarly,  from (\ref{detConn2}) one finds that the determined (trace) part in the ${\bf 56}$ reduces as, 
 \be
 \begin{split}
  \big\langle {\cal Q}_{\underline{B}\underline{A}}{}^{\,\underline{B}}\big\rangle  
  \ = \ ({\cal V}^{-1})_{\underline{A}}{}^{\,{N}}\big(\partial_M(U^{-1})_{{N}}{}^{\,M}
  -3\,\rho^{-1}(U^{-1})_{{N}}{}^{\,M}\partial_M\rho\big)
  \ = \ 2\,\rho\,({\cal V}^{-1})_{\underline{A}}{}^{\,{N}}\,\vartheta_{{N}}\;, 
 \end{split} 
 \ee 
identifying it with the T-tensor corresponding to the trombone embedding tensor. 
Thus, the parts of the SU$(8)$ connection that are determined geometrically 
by generalized torsion and metricity constraints,  
upon Scherk-Schwarz reduction naturally identify with the T-tensor. 
Comparing with the definitions of \cite{deWit:2007mt,LeDiffon:2011wt}, the relation 
is explicitly given by  
 \bea\label{SSTrel}
  \big\langle \,\big[\,{\cal Q}_{\underline{A}\underline{B}}{}^{\,\underline{C}}\,\big]_{\rm 912}\,\big\rangle ~=~ 
  \frac{2}{7}\,\rho\, T_{\underline{A}\underline{B}}{}^{\,\underline{C}} \;,
\qquad
  \big\langle {\cal Q}_{\underline{B}\underline{A}}{}^{\underline{B}}\big\rangle  ~=~ 
   -2\,\rho\,T_{\underline{A}}\;.
\eea
These relations makes the following comparison with gauged supergravity 
straightforward.

In order to perform that comparison in detail we note that the 
${\bf 912}$ representation of the T-tensor decomposes under SU$(8)$ as 
 \be
  T_{\underline{A}\underline{B}}{}^{\,\underline{C}}\; : \qquad
  {\bf 912} \ \rightarrow \ {\bf 420} \ + \  {\bf 36} \ + \  {\rm c.c.}\;, 
 \ee 
which implies, for instance, that the component $T^{ij}{}_{kl}{}^{pq}$ of $T_{\underline{A}\underline{B}}{}^{\,\underline{C}}$
can be parametrized as 
 \be
   T^{ij}{}_{kl}{}^{pq} \ = \ \frac{2}{3}\delta_{[k}{}^{[p} \, T_{l]}{}^{q]ij}\;, 
 \ee  
in terms of a tensor $T_{i}{}^{jkl}$, which in turn can be decomposed as 
  \be\label{A1A2DEF}
  T_{i}{}^{jkl} \ = \ -\frac{3}{4}{A}_{2i}{}^{jkl}-\frac{3}{2}{A}_{1}{}^{j[k}\,\delta^{l]}{}_{i}\;. 
 \ee 
Here $A_1{}^{ij}$ is symmetric  and hence lives in the ${\bf 36}$, 
and $A_2$ satisfies $A_{2i}{}^{jkl}=A_{2i}{}^{[jkl]}$, $A_{2i}{}^{ikl}=0$, 
and hence lives in the ${\bf 420}$. The tensors $A_1$ and $A_2$ 
thus defined contribute to the fermion shifts in the supersymmetry variations 
and the scalar potential of gauged supergravity. 
Similarly, the trombone T-tensor, i.e., the flattening of $\vartheta_{M}$, decomposes 
as 
 \be\label{Ttrombone}
  T_{\underline{A}} \ \equiv \ ({\cal V}^{-1})_{\underline{A}}{}^{M}\vartheta_M  
  \ = \ \big( \,T_{ij}\,,\; T^{ij}\,) \ \equiv \ \big( \,B_{ij}\,,\; B^{ij}\,)\;, 
 \ee  
in terms of an antisymmetric tensor $B_{ij}$ (and its complex conjugate). 
Via eqs.~(\ref{A1A2DEF}) and (\ref{Ttrombone}) and the relations (\ref{SSTrel}) 
for the Scherk-Schwarz reduction of the connections we can express  the latter in terms 
of the gauged supergravity tensors. There is the following subtlety, however: 
The components of the SU$(8)$ connection entering, say, the fermionic
supersymmetry variations are the ${\cal Q}_{Mi}{}^{j}$ from (\ref{Qij}), while the ${\bf 912}$ 
projection in (\ref{SSTrel}) shuffles its components around. 
A slightly technical group-theoretical analysis, whose details we defer to 
the appendix, shows that the net effect is a rescaling of the $A_2$ contribution in 
the T-tensor by $\frac73$ (see (\ref{Q2}) in the appendix), while 
$A_1$ is untouched. 
We then find  
\bea
\big\langle \,\big[\, {\cal Q}^{ij}{}_{km}{}^{ln}\,\big]\,\big\rangle 
 &=& 
\frac27\,\rho\,\left(
-\frac12\cdot\frac73\,\delta_{[k}{}^{[l} A_{2m]}{}^{n]ij}-\delta_{km}{}^{[i[l} A_1{}^{n]j]}
\right)
\nonumber\\
&&{}-
\frac{16}{27}\,\rho\,\delta_{km}{}^{[i[l} B^{n]j]} - \frac{2}{27}\,\rho\, \delta_{km}{}^{ln}\,B^{ij}  \;, 
\label{QQ}
\eea
where the square bracket indicates projection onto the determined part of 
the connections. 
In particular, we obtain 
\bea\label{QREsults}
 \big\langle {\cal Q}^{ij}{}_{j}{}^{k}\big\rangle &=&
-\rho\,\left(
A_1{}^{ik}
-2\,B^{ik}
\right)
\;,\nonumber\\
 \big\langle {\cal Q}^{[ij}{}_{k}{}^{l]}\big\rangle &=&
-\frac13\, \rho\,\left(
  A_{2k}{}^{lij}
+2\,\delta_{k}{}^{[i} B^{jl]}
\right)
\;,
\eea
for the trace and the total antisymmetrization of ${\cal Q}_{\underline{A}}{\,}_i{}^j$ from (\ref{Qij}).

With this, we can turn to the Scherk-Schwarz reduction of the fermionic sector.
The Scherk-Schwarz ansatz for the fermions is simply governed 
by their respective density weights, 
 \be
  \psi_{\mu}{}^{i}(x,Y) \ = \  \rho^{-\frac{1}{2}}(Y)\,\psi_{\mu}{}^{i}(x)\;, \qquad
  \chi^{ijk}(x,Y) \ = \ \rho^{\frac{1}{2}}(Y)\chi^{ijk}(x)\;,
 \ee 
in accordance with their behavior under generalized diffeomorphisms.\footnote{
We note, in particular, that the ansatz does not carry the Killing spinors of the internal manifold
as one might have expected in analogy to standard Kaluza-Klein compactifications. 
This appears natural, since for the general class of reductions to be discussed
the internal space may not even possess sufficiently many Killing spinors.
With respect to the supersymmetric reduction ansatz of \cite{de Wit:1986iy} for the $S^7$ case
this corresponds to a different ${\rm SU}(8)$ gauge choice.} 
Indeed, we will confirm in the following that this ansatz reproduces precisely 
the supersymmetry variations of gauged supergravity. 
Consider first the gravitino variation in (\ref{fullSUSY}), 
  \be
   \big\langle \,\delta_{\epsilon}\psi_{\mu}{}^{i}\,\big\rangle \ = \ 
   \big\langle2\,{\cal D}_{\mu}\epsilon^i \big\rangle  -4\, 
    \big\langle\, \widehat{\nabla}^{ij}\big(\gamma_{\mu}\epsilon_j\big)\,\big\rangle\;.
  \ee  
By the results of the previous section the first term reduces to the correct 
gauge covariantized external derivative ${\cal D}_{\mu}\epsilon^i$ of gauged supergravity.
For the second term using that $\gamma_{\mu}\epsilon_j$ has weight~$\tfrac{3}{4}$ the covariant 
derivative above reads 
 \bea
  \widehat{\nabla}^{ij}\big(\gamma_{\mu}\epsilon_j\big) &=& ({\cal V}^{-1})^{M\,ij} \Big(e^{\frac{3}{8}} \partial_M
  \big(e^{-\frac{3}{8}}\gamma_{\mu}\epsilon_j\big) + \frac{1}{4}e^{\mu\alpha}\partial_M e_{\mu}{}^{\beta}\,
  \gamma_{\alpha\beta}\gamma_{\mu}\epsilon_j\Big)+\frac{1}{2}Q^{ij}{}_{ j}{}^{k}\gamma_{\mu}\epsilon_k
  \nonumber\\
  &&{}
  -\frac1{16}\, {\cal F}_{\nu\rho}{}^{ij}\,\gamma^{\nu\rho}\gamma_\mu\,\epsilon_j
  \;,
 \eea
 with the flattened field strength from (\ref{YM}).
Upon inserting the Scherk-Schwarz ansatz the terms in the first parenthesis are 
actually $Y$-independent, so this derivative vanishes. Also the second term is zero because under 
Scherk-Schwarz reduction the first 
factor is proportional to $\eta^{\alpha\beta}$, which vanishes upon contraction with $\gamma_{\alpha\beta}$. 
Together, we find with (\ref{QREsults}) 
 \be
  \big\langle \,\delta_{\epsilon}\psi_{\mu}{}^{i}\,\big\rangle \ = \ 
  \rho^{-1/2} \left(
  2\,{\cal D}_\mu \epsilon^i 
  +\frac14\,  {\cal F}_{\nu\rho}{}^{ij}\,\gamma^{\nu\rho}\gamma_\mu\,\epsilon_j
   +2\, (A_{1}{}^{ij}-2B^{ij})\,\gamma_{\mu}\epsilon_j \right)
 \;.
 \ee 
These are precisely the gravitino variations of gauged supergravity including trombone 
gaugings, as given in \cite{LeDiffon:2011wt} (taking into account the 
change of normalization explained in footnote~\ref{foot:norm}).
In complete analogy, we obtain with the second relation from (\ref{QREsults})
 \bea
 \big\langle\,  \delta_{\epsilon}\chi^{ijk}\, \big\rangle &=& 
 \rho^{1/2}\left(
 -2 \sqrt{2} {{\cal P}}_{\mu}{}^{ijkl} \gamma^{\mu} \epsilon_l 
 + \frac{3\sqrt{2}}{4} {{\cal F}}_{\mu \nu}{}^{[ij} \gamma^{\mu \nu} \epsilon^{k]} 
 -2\sqrt{2}\, A_{2\,l}{}^{ijk} \epsilon^l - 4\sqrt{2}\,B^{[ij} \epsilon^{k]}
 \right)
 \;.\nonumber\\
 \eea 
Again, this is precisely the expected result for the fermion supersymmetry variation 
in gauged supergravity including trombone gauging.  

Finally, we turn to the Scherk-Schwarz reduction of the scalar potential given in (\ref{scalarPot}), 
with the generalized scalar curvature defined implicitly by (\ref{covPOT}). 
Upon Scherk-Schwarz reduction the latter equation reads 
 \be\label{calRDEF}
  \frac{1}{16}\big\langle {\cal R}\epsilon_i\big\rangle \ = \  
  6\,\big\langle \nabla_{ik}\nabla^{kj}\epsilon_j\big\rangle
  +3\,\big\langle \nabla^{kj}\nabla_{[ik}\epsilon_{j]}\big\rangle\;,
 \ee 
since $\langle\nabla_M g_{\mu\nu}\rangle =0 $\,. 
It is then straightforward to determine the Scherk-Schwarz reduction of the various terms 
in the potential by successive action of (\ref{QQ}). For instance, the first term reads 
 \be
  \big\langle \nabla_{ik}\nabla^{kj}\epsilon_j\big\rangle \ = \  
    \big\langle e^{-\frac{1}{8}}\partial_{ik}\big(e^{\frac{1}{8}}\nabla^{kj}\epsilon_j\big)
    -\frac{1}{2}Q_{ik\,l}{}^{k}\nabla^{lj}\epsilon_j\big\rangle 
    \ = \   -\frac{1}{2} \big\langle Q_{ik\,l}{}^{k}\big\rangle \big\langle \nabla^{lj}\epsilon_j\big\rangle \;, 
 \ee 
where the vanishing of the first term in here follows from $\nabla\epsilon$ having weight $-\frac{1}{4}$, 
for which all $Y$-dependence cancels. Similarly, working out the inner covariant derivative 
we find 
 \be    
  \big\langle \nabla_{ik}\nabla^{kj}\epsilon_j\big\rangle \ = \ -\frac{1}{4} \big\langle Q_{ik\,l}{}^{k}\big\rangle
  \big\langle
  Q^{lj}{}_{j}{}^{k} \big\rangle \big\langle \epsilon_k\big\rangle\;, 
 \ee   
which by use of (\ref{QREsults}) can be expressed in terms of the gauged supergravity 
quantities $A_1$, $A_2$ and $B$. Performing the same computation for the final 
two terms in (\ref{calRDEF}) we obtain eventually
 \bea
  \frac{1}{16}\big\langle {\cal R}\epsilon_i\big\rangle  &=&
\rho^{3/2}\,
\Big(\,\frac32\,A_{im}A^{jm} -\frac1{12}A_{i}{}^{kmn} A^j{}_{kmn}
+3A_{im}B^{jm}-3 B_{im}A^{jm} 
\nonumber\\
&&{}
\;\;-\frac16\, A_i{}^{jmn} A_{mn}
+\frac56\,A^j{}_{imn}B^{mn} -\frac{64}9\,B_{im}B^{jm} +\frac59\,B_{mn}B^{mn} \delta_i^j\,
\Big) \epsilon_j
 \nonumber\\
 &=&
 \rho^{3/2}\,\Big(\, \frac3{16}\,A_{mn}A^{mn}-\frac1{96}A_{l}{}^{kmn} A^l{}_{kmn} -\frac13\,B_{mn}B^{mn}\, \Big)\,\epsilon_i
 \;,
 \eea 
where we have used in the last equation the quadratic constraints satisfied by the embedding tensor, 
c.f.~(5.3)--(5.5) of \cite{LeDiffon:2011wt}.\footnote{We
recall that these constraints are automatically satisfied as a consequence of the section constraints~(\ref{sectioncondition}).}
This gives precisely the correct scalar potential of gauged supergravity
(or, more precisely, the correct contribution to the Einstein field equations in the presence of trombone 
gauging $B_{ij}\neq0$). 
An equivalent calculation for the scalar potential has been done in~\cite{Aldazabal:2013mya} via the generalized 
Ricci tensor of~\cite{Coimbra:2011ky} with the full match to gauged supergravity in absence of the trombone parameter.

To summarize, we have shown that the generalized Scherk-Schwarz ansatz is consistent in the full 
exceptional field theory and exactly reproduces all field equations and transformation rules
of the lower-dimensional gauged supergravity, provided the twist matrices satisfy (\ref{consistent_twist})
and the section condition (\ref{sectioncondition}).

\section{Sphere and hyperboloid compactifications}
\label{sec:sphere}

In this section we construct explicit solutions to the twist equations (\ref{consistent_general}) within the
${\rm SL}(5)$, ${\rm E}_{6(6)}$, and ${\rm E}_{7(7)}$ exceptional field theories. The twist matrices live within
a maximal ${\rm SL}(n)$ subgroup (for $n=5$, $n=6$, and $n=8$, respectively), and describe
consistent truncations to lower-dimensional theories with gauge group ${\rm CSO}(p,q,r)$ for $p+q+r=n$\,.
For $r=0$, the corresponding internal spaces are warped hyperboloids and spheres, for 
$r>0$ they also include factors of warped tori.

\subsection{${\rm SL}(n)$ twist equations}

To begin with, let us study the case of the $D=7$, ${\rm SL}(5)$ theory.  As it turns out, this case 
already exhibits all the structures relevant  for the general sphere and hyperboloid
compactifications. 
Although the full ${\rm SL}(5)$ \eft (including $D=7$ metric, vector, and $p$-form fields) 
has not yet been constructed, the internal (scalar) sector has been studied 
in some detail~\cite{Berman:2010is,Coimbra:2011ky,Berman:2012uy,Park:2013gaj}.
In this case, the underlying group is ${\rm SL}(5)$ and vector fields ${\cal A}_\mu{}^{AB}$ and internal coordinates
$Y^{AB}$ transform in the antisymmetric ${\bf 10}$ representation, i.e.
\bea
{\cal A}_\mu{}^{AB} &=& {\cal A}_\mu{}^{[AB]}\;,
\qquad
Y{}^{AB} ~=~Y{}^{[AB]}\;,
\qquad
A, B = 0, \dots, n-1
\;,
\eea
with $n=5$\,. In order to prepare 
the ground for the general case, 
we will in the following keep $n$ 
arbitrary and only in the final formulas 
specify to $n=5$\,.\footnote{
It is in this `${\rm SL}(n)$ generalized geometry' that the generalized parallelizability
of spheres has been discussed in~\cite{Lee:2014mla}.}
The section conditions in this case take the ${\rm SL}(n)$ covariant form
\bea
\partial_{[AB} \otimes \partial_{CD]} &\equiv& 0
\;.
\label{section_SL}
\eea
The reduction ansatz for the vector field is given by (\ref{SSeAB})
\bea
{\cal A}_\mu{}^{AB}(x,Y) &=& \rho^{-1}\,(U^{-1})_{CD}{}^{AB} \, A_\mu{}^{CD}(x)
~=~ \rho^{-1}\,(U^{-1})_C{}^A (U^{-1})_D{}^B\,  A_\mu{}^{CD}(x)
\;,
\eea
in terms of an ${\rm SL}(n)$-valued $n\times n$ twist matrix $U_A{}^B$\,.
For $D=7$ maximal supergravity, the embedding tensor resides in the 
$[2,0,0,0] \oplus [0,0,1,1]$ representation~\cite{Samtleben:2005bp}
\bea
\Theta_{AB,C}{}^{D}&=&
\delta^{D}_{[A}\,\eta^{\phantom{D}}_{B]C}
+ Z_{ABC}{}^D  
\;,
\label{linear}
\\
&&{}\mbox{with}\quad
\eta_{AB} = \eta_{(AB)}\;,\quad
Z_{ABC}{}^D = Z_{[ABC]}{}^D\;,\quad Z_{ABC}{}^C = 0
\;.
\nonumber
\eea
Accordingly, the twist equations (\ref{consistent_general}) take the form
\bea
\left[{\cal X}_{AB,C}{}^D\right]_{(\mathbb{P})} &\stackrel{!}{=}&
\rho \,\left(\delta^{D}_{[A}\,\eta^{\phantom{D}}_{B]C}
+ Z_{ABC}{}^D\right)  
\;,
\nonumber\\
  \partial_{CD} (U^{-1})_{AB}{}^{CD}  
- (D-1)\,\rho^{-1}\partial_{CD} \rho  \, (U^{-1})_{AB}{}^{CD}  &\stackrel{!}{=}& 
\rho\,(D-2)\,\vartheta_{AB}
\;,
\label{consistent_SL}
\eea
with $D=7$, which for the purpose of later generalisations 
we also keep arbitrary for the moment and only specify in the final formulas.
Here, ${\cal X}_{AB,C}{}^D$ denotes the ${\rm SL}(n)$ version of (\ref{XXX})
\bea
{\cal X}_{AB,C}{}^D&\equiv& (U^{-1})_{AB}{}^{GH} (U^{-1})_C{}^E \, \partial_{GH} U_E{}^D
\;,
\label{XX}
\eea
and  the projection $[\cdot]_{({\mathbb{P}})}$ refers to the 
projection onto the representations of $\eta_{AB}$ and $Z_{ABC}{}^D$ from (\ref{linear}). We can
thus write the first equation of (\ref{consistent_SL}) more explicitly as
\bea
\partial_{CD} (U^{-1})_{(A}{}^C (U^{-1})_{B)}{}^{D}  &\stackrel{!}{=}&
\frac12\,(1-n)\, \rho \,\eta_{AB}
\;,\nonumber\\
(U^{-1})_{ABC}{}^{GHE}  \, \partial_{GH} U_E{}^D -\frac1{(n-2)}\,
\partial_{GH} (U^{-1})_{[AB}{}^{GH}\,   \delta_{C]}{}^D
&\stackrel{!}{=}&
\rho \, Z_{ABC}{}^D  
\;.
\label{consistent_SL0}
\eea
For later use, let us record that in terms of irreducible ${\rm SL}(n)$ representations (for general value of $n$) 
the consistency equations (\ref{consistent_SL}), (\ref{consistent_SL0}) constrain the
\bea{}
[0,1,0,\dots,0] ~\oplus~ [2,0,0,\dots,0] ~\oplus~ [0,0,1,\dots,0,1]
\;, 
\label{conNcon}
\eea
part of the ${\cal X}_{AB,C}{}^D$,
but leave its
\bea
[1,1,0, \dots,0,1]
\;,
\label{conNnot}
\eea
part unconstrained. For the present case $n=5$, this translates into a constrained part 
${\bf 10}\oplus{\bf 15}\oplus{\bf 40'}$ and an unconstrained ${\bf 175}$\,.

In the following we will construct the twist matrices corresponding to sphere and hyperboloid compactifications.
These satisfy the additional conditions
\bea
Z_{ABC}{}^D &=& 0\;,\qquad \vartheta_{AB} ~=~ 0\;.
\label{ZT}
\eea
This is consistent with the fact that the resulting gauged supergravities are described by an embedding tensor 
$\eta_{AB}$ and no trombone symmetry is excited in these compactifications.
The consistency equations (\ref{consistent_SL}), (\ref{consistent_SL0}) thus take the stronger form
\begin{subequations}
\bea
 \partial_{CD} (U^{-1})_{(A}{}^C (U^{-1})_{B)}{}^{D}  &\stackrel{!}{=}&
\frac12\,(1-n)\, \rho \,\eta_{AB}\;,
\label{con1}\\
 \partial_{CD} \Big[ \rho^{1-D}\, (U^{-1})_{AB}{}^{CD}  \Big]
&\stackrel{!}{=} & 0
\;,
\label{con2}\\
(U^{-1})_{ABC}{}^{GHE}  \, \partial_{GH} U_E{}^D &\stackrel{!}{=}& \frac{D-1}{n-2}\,
 (U^{-1})_{[AB}{}^{GH}\,   \delta_{C]}{}^D\,  \rho^{-1}  \partial_{GH} \rho \,  
\;.
\label{con3}
\eea
\end{subequations}
In the following, we will construct solutions to these equations for arbitrary constant $\eta_{AB}$\,.
Let us note that the ${\rm SL}(n)$ covariance of these consistency equations allows to bring
$\eta_{AB}$ into diagonal form 
\begin{align}
  \eta_{AB} &= 
  {\rm diag}\,(\,\underbrace{1, \dots,}_{p}\underbrace{-1,\dots,}_{q} \underbrace{0, \dots}_{r}\,) \;,
  \qquad
  \mbox{with}\;\; p+q+r=n\;,
  \label{YinCSO}
\end{align}
upon conjugation of $U$ and constant ${\rm SL}(n)$ rotation of the internal coordinates.
The resulting reduced theories are gauged supergravities with gauge group ${\rm CSO}(p,q,r)$,
defined as the ${\rm SL}(n)$ subgroup preserving (\ref{YinCSO}). For $r=0$, this is the non-compact semisimple group ${\rm SO}(p,q)$,
for $r>0$ it corresponds to the non-semisimple group with algebra spanned by matrices $T_{AB}$:
\bea
(T_{AB})_C{}^D&\equiv&  \eta_{C[A}\,\delta_{B]}{}^D
\;.
\eea

\subsection{Sphere and hyperboloid solutions}
\label{subsec:sphere}

Recall that the twist matrices $U_A{}^B$ are not only subject to the consistency conditions (\ref{con1})--(\ref{con3}),
but also to the section conditions (\ref{section_SL}). In order to solve the latter, we make the following ansatz
\bea
\partial_{ij} U_A{}^B &=& 0
\;,
\qquad \mbox{ for}\quad i,j = 1, \dots, n-1
\;,
\label{sectionsol_SL}
\eea
i.e.\ we restrict the coordinate dependence of $U_A{}^B$ to the $(n-1)$ coordinates $y^i \equiv Y^{0i}$\,.
For the ${\rm SL}(5)$ theory this corresponds to restricting the internal part of the exceptional space-time from 10
to the 4 coordinates that extend the $D=7$ theory to eleven-dimensional supergravity~\cite{Berman:2010is}.
After this reduction, the first of the twist equations (\ref{con1}) splits into the pair of equations
\begin{align}
 \partial_{i} (U^{-1})_{0}{}^0 (U^{-1})_{m}{}^{i} 
  -   \partial_{i} (U^{-1})_{m}{}^i (U^{-1})_{0}{}^{0} 
 &=
 \partial_{i} (U^{-1})_{0}{}^i (U^{-1})_{m}{}^{0} 
-  \partial_{i} (U^{-1})_{m}{}^0 (U^{-1})_{0}{}^{i}
\;,
\label{twist_exp}\\
 \partial_{i} (U^{-1})_{(m}{}^0 (U^{-1})_{n)}{}^{i} -   \partial_{i} (U^{-1})_{(m}{}^i (U^{-1})_{n)}{}^{0} &=
\left(\partial_{i} (U^{-1})_{0}{}^0 (U^{-1})_{0}{}^{i} -   \partial_{i} (U^{-1})_{0}{}^i (U^{-1})_{0}{}^{0} \right)
 \,\eta_{mn}
\;,
\nonumber
\end{align}
while the density factor $\rho$ is obtained from the $(AB)=(00)$ component as
\bea
\rho&=& \frac{1-n}{2} \left( \partial_{i} (U^{-1})_{0}{}^0 (U^{-1})_{0}{}^{i} -   \partial_{i} (U^{-1})_{0}{}^i (U^{-1})_{0}{}^{0} 
\right)
\;.
\label{readrho}
\eea
Here $\eta_{mn}$ is the reduction of $\eta_{AB}$ (\ref{YinCSO}) to the last $n-1$ coordinates, i.e.\ the diagonal matrix
\begin{align}
  \eta_{mn} &= 
  {\rm diag}\,(\,\underbrace{1, \dots,}_{p-1}\underbrace{-1,\dots,}_{q} \underbrace{0, \dots}_{r}\,) 
\;.
\label{etapqr}
\end{align}
We will first treat the case $r=0$ of non-degenerate $\eta_{mn}$ and subsequently extend the 
discussion to the general case.

\subsubsection{The case ${\rm SO}(p,q)$}

For $\eta_{mn}$ given by (\ref{etapqr}) with $r=0$, equations (\ref{twist_exp}) can be solved by the following explicit 
${\rm SL}(n)$ ansatz
\bea
\label{Upq}
(U^{-1})_0{}^0&\equiv&  (1-v)^{(n-1)/n}\;,
\nonumber\\
(U^{-1})_0{}^i &\equiv& \eta_{ij} y^j\,(1-v)^{(n-2)/(2n)}\,K(u,v)\;,
\nonumber\\
(U^{-1})_i{}^0 &\equiv& \eta_{ij} y^j\,\,(1-v)^{(n-2)/(2n)}\;,
\nonumber\\
(U^{-1})_i{}^j &\equiv& (1-v)^{-1/n}\left( \delta^{ij}\,+  \eta_{ik} \eta_{jl} \, y^k  y^l\,K(u,v)\right)\;,
\eea
with $n=p+q$ and the combinations 
$u\equiv y^iy^i$, $v\equiv y^i\eta_{ij}y^j$. Upon inserting this ansatz into (\ref{twist_exp}), these equations reduce
to a single differential equation for the function $K(u,v)$, given by
\bea
2(1-v)\left(u \,\partial_v K+v\,\partial_u K \right)&=&
\left((1+q-p)(1-v) -u\right) K -1
\;.
\label{diffK}
\eea
This equation takes a slightly simpler form upon change of variables
\bea
u\equiv r^2\,{\rm cosh}\,\varphi\;,\qquad v\equiv r^2\,{\rm sinh}\,\varphi
\;,
\label{uvr}
\eea
after which it becomes an ordinary differential equation in $\varphi$
\bea
2\left(1-r^2\,{\rm sinh}\varphi\right) \partial_\varphi K &=&
\left((1+q-p)(1-r^2\,{\rm sinh}\,\varphi)-r^2\,{\rm cosh}\,\varphi \right) K -1
\;.
\eea
This can be solved analytically for any pair of integers $(p,q)$. 
We have to treat separately the cases $q=0$ and $p=1$ (corresponding to ${\rm SO}(p)$ and ${\rm SO}(1,q)$ gaugings, respectively)
for which $u=\pm v$ and the change of variables (\ref{uvr}) does not make sense.
In the former case $u=v$, and equation (\ref{diffK}) reduces to
\bea
2\,u\,(1-u)K'&=&
\left( 1-p+pu-2u  \right) K -1
\;,
\eea
with the particular solution
\bea
K &=& -{}_2F_1[1,(p-2)/2; 1/2;1-u]
\;,
\eea
in terms of the hypergeometric function $_2F_1$\,.
Similarly, for $p=1$, we have $u=-v$, and equation (\ref{diffK}) reduces to
\bea
-2\,u \,(1+u) K'&=&
\left(q+q u -u \right) K -1
\;,
\eea
with particular solution
\bea
K &=& {}q^{-1} {}(1+u(1-q)\, _2F_1[1,(1+q)/2;1/2;1+u])
\;.
\eea
Finally, the density factor $\rho$ can be read off from (\ref{readrho}) as
\bea
\rho~=~\rho_{p,q} &\equiv& (1-v)^{(n-4)/(2n)}
\;.
\label{rhopq}
\eea
We have thus fully determined the twist matrix $U$ and the density factor $\rho$ 
and entirely solved the first of the twist equations (\ref{con1}). It remains to verify the other two equations.

With the twist matrix given by (\ref{Upq}) and using the differential equation 
(\ref{diffK}) for the function $K$, it is straightforward to verify that
\bea
\partial_i \left[ (U^{-1})_{[A}{}^0 (U^{-1})_{B]}{}^i \right] &=&
\frac{n-2}{n}\,\frac{2\,\eta_{ij}y^j}{1-v}
\,
 (U^{-1})_{[A}{}^0 (U^{-1})_{B]}{}^i 
\;.
\label{con2_A}
\eea
Together with the form of the density factor $\rho$ from (\ref{rhopq}), we thus find, that the second equation 
(\ref{con2}) is identically satisfied provided we have the relation
\bea
\frac12\,(D-1) &=& \frac{n-2}{n-4}
\;,
\label{ddd}
\eea
relating the number of external space-time dimensions to the size of the group ${\rm SL}(n)$\,.
Fortunately, this relation holds precisely in the case $D=7$, $n=5$ that we are interested in. 
We have thus shown that also the second twist equation (\ref{con2}) holds for our choice of
twist matrix and density factor.
Let us note that integer solutions of equation (\ref{ddd}) are very rare and essentially restrict to 
\bea
(D,n)=(7,5)\;,\qquad
(D,n)=(5,6)\;,\qquad
(D,n)=(4,8)\;,
\label{457}
\eea
in which we recover the dimensions of the known sphere compactifications AdS$_D \times {\rm S}^{n-1}$
of eleven-dimensional and type IIB supergravity. We will come back to this in section~\ref{subsec:E6E7}.
As a last consistency check, one verifies by direct computation that the last twist equation (\ref{con3})
is also identically satisfied for (\ref{Upq}) with (\ref{diffK}).
This essentially follows from the fact that no object fully antisymmetric  in three indices $[ABC]$ can be 
constructed from the explicit $\eta_{ij} y^j$\,.

To summarize, we have shown that the ${\rm SL}(n)$ 
twist matrix $U$ given by (\ref{Upq}) with (\ref{diffK}), together with the density factor (\ref{rhopq}) 
satisfies the consistency equations (\ref{con1})--(\ref{con3}), 
provided the integer relation (\ref{ddd}) holds. In the next section, we
will generalize this solution to include the case $r>0$.

\subsubsection{The case ${\rm CSO}(p,q,r)$}

The solution of equations equations (\ref{con1})--(\ref{con3}) derived in the previous section can be generalized
to the case $r>0$, in which the reduced theory comes with the gauge group ${\rm CSO}(p,q,r)$.
A natural ansatz for the ${\rm SL}(p+q+r)$ twist matrix in this case is given by
\bea
(U^{-1})_A{}^B &=& \left(
\begin{array}{c:c}
\beta^{-r}\,
U^{-1}_{(p,q)}& 0 
\\ \hdashline
0 & \beta^{p+q}\,\mathbb{I}_r  \end{array}
\right)
\;,
\label{Upqr}
\eea
where $U^{-1}_{(p,q)}$ denotes the ${\rm SO}(p,q)$ solution from (\ref{Upq}), (\ref{diffK}), and 
$\mathbb{I}_r$ is the $r\times r$ identity matrix. The factor $\beta=\beta(v)$ is a function of $v\equiv y^i\eta_{ij} y^j$
and put such that the determinant of the twist matrix remains equal to one.
Note that the twist matrix only depends on coordinates $y^i$, $i=1, \dots, p+q-1$\,.

Let us now work out equations (\ref{con1})--(\ref{con3}) for the ansatz (\ref{Upqr}).
The first equation (\ref{con1}) is solved identically by this ansatz without any assumption on the function $\beta$,
as a mere consequence of the fact that $U_{(p,q)}$ solves the corresponding equations for ${\rm SO}(p,q)$. Indeed, all components
of this equation in which one of the free indices $(AB)$ takes 
values beyond $p+q-1$, hold trivially due to the block-diagonal
structure of the twist matrix (\ref{Upqr}) and the fact that the matrix does not depend on the last $r$ coordinates.
The constraint equations then simply reduce to their ${\rm SO}(p,q)$ analogues. The presence of the factor $\beta^{-r}$ does not 
spoil the validity of the equation, but contributes to the density factor $\rho$ as
\bea
\rho &=& \rho_{p,q,r} ~\equiv~ \beta^{-2r}\,\rho_{p,q}
\;,
\label{rhopqr}
\eea
with $\rho_{p,q}$ from (\ref{rhopq}).
We continue with the second equation (\ref{con2}) which now takes the form
\bea
 \partial_{CD} \Big[ \rho_{p,q}^{1-D}\,\beta^{2r\,(D-2)}\,  
 (U_{(p,q)}^{-1})_{AB}{}^{CD}  \Big]
&\stackrel{!}{=} & 0
\;,
\eea
and thus reduces to the identity (\ref{con2_A}) provided we choose $\beta$ such that
\bea
\rho_{p,q}^{1-D}\,\beta^{2r\,(D-2)} &=& 
(1-v)^{-(p+q-2)/(p+q)}
\;.
\eea
With (\ref{rhopq}), this reduces to
\bea
\beta^r &=& (1-v)^{(D-3)/(4(D-2))-1/(p+q)}
\;.
\label{beta0}
\eea
Even though the ${\rm CSO}(p,q,r)$ case seems to admit more freedom in that we are not bound by a relation (\ref{ddd})
to fix the size of the external space-time, we will for the moment restrict to the three principal cases
(\ref{457}), i.e.\ keep the additional relation (\ref{ddd})
\bea
\frac12\,(D-1) &=& \frac{p+q+r-2}{p+q+r-4}
\;,
\label{dddpqr}
\eea
and describe reductions to four, five and seven dimensions, respectively.
Then, the form of the scale factor $\beta$ from (\ref{beta0}) simplifies to
\bea
\beta &=& (1-v)^{-1/((p+q)(p+q+r))}
\;.
\label{beta}
\eea
Together with (\ref{Upqr}) this fully defines the twist matrix that solves the consistency equations 
(\ref{con1})--(\ref{con3})  for $\eta_{AB}$ from (\ref{YinCSO}). As above, the last equation
(\ref{con3}) is verified by explicit calculation.

To summarize, we have shown that the ${\rm SL}(p+q+r)$ 
twist matrix $U_M{}^N$ given by (\ref{Upq}), (\ref{Upqr}), (\ref{beta}), together with the density factor (\ref{rhopqr}),
satisfies the consistency equations (\ref{con1})--(\ref{con3}) for the general $\eta_{AB}$ of (\ref{YinCSO}), 
provided the integer relation (\ref{dddpqr}) holds.
By the above discussion, this applies in particular to the case $D=7$, ${\rm G}={\rm SL}(5)$, 
and implies that the resulting twist ansatz describes a consistent truncation
of the corresponding \eft. Since this twist matrix also falls into the class (\ref{sectionsol_SL}) of solutions
to the section conditions, this generalized Scherk-Schwarz ansatz describes consistent truncations
of the 
full $D=11$ supergravity down to seven-dimensional supergravities with gauge groups ${\rm CSO}(p,q,r)$, ($p+q+r=5$). 
We will work out in section~\ref{subsec:induced} the internal background metrics induced by these twist matrices,
in order to get the geometrical perspective for these compactifications.

\subsection{${\rm E}_{6(6)}$ and ${\rm E}_{7(7)}$ twist equations}
\label{subsec:E6E7}

In this section, we show that the ${\rm SL}(n)$ twist matrices found in the previous section can also be used for the
construction of solutions to the consistency equations (\ref{consistent_general}) in the exceptional field theories 
with groups ${\rm E}_{6(6)}$ and ${\rm E}_{7(7)}$. 
The structure underlying this construction is the embedding of the ${\rm SL}(n)$ twist matrices via
\bea
{\rm SL}(6) &\subset& {\rm E}_{6(6)}\;,\qquad
{\rm SL}(8) ~\subset~ {\rm E}_{7(7)}\;,
\label{SLinE}
\eea
respectively, inducing a decomposition of the  ${\rm E}_{n(n)}$ coordinates according to
\bea
Y^M &\longrightarrow& \left\{ Y^{[AB]},\;\dots \right\} 
~\longrightarrow~ \left\{ y^i,\;\dots \right\} \;,
\qquad \mbox{with}\quad y^i \equiv Y^{[0i]}
\;,
\nonumber\\
&&{}
A, B = 0, \dots, n-1\;,\qquad i=1, \dots, n-1
\;.
\label{SLinE1}
\eea
Together with a solution of the section constraint achieved by restricting the coordinate dependence
of all fields according to
\bea
\Phi(Y^M) &\longrightarrow& \Phi(y^i)
\;,
\label{SLinE2}
\eea
and the fact that both cases (\ref{SLinE}) correspond to a solution of the integer relation (\ref{ddd})
(with $(D,n)=(5,6)$ and $(D,n)=(4,8)$, respectively), this structure turns out to be sufficient to ensure that
the ${\rm SL}(n)$ twist matrices constructed above also define solutions to the consistency equations 
(\ref{consistent_general}) of these larger exceptional field theories.
The corresponding Scherk-Schwarz ansatz then defines lower-dimensional theories with
embedding tensor describing the gauge groups ${\rm CSO}(p,q,r)$, ($p+q+r=n$).

\subsubsection{${\rm E}_{6(6)}$}

For details about the ${\rm E}_{6(6)}$ exceptional field theory, we refer to \cite{Hohm:2013vpa}.
It is formulated on an internal space of 27 coordinates $Y^M$ in the fundamental representation of ${\rm E}_{6(6)}$,
with the section condition given by the 27 equations $d^{KMN} \partial_M \otimes \partial_N \equiv 0$, with the cubic invariant $d^{KMN}$\,.
In this case, the subgroup ${\rm SL}(6)$ is embedded into ${\rm E}_{6(6)}$ via 
\bea
{\rm SL}(6)~\subset~   {\rm SL}(6)\times {\rm SL}(2)~\subset~ {\rm E}_{6(6)}
\label{662}
\eea
with the internal coordinates decomposing as
\bea
\bar{\bf 27} &\longrightarrow& (15,1)+(6',2)
\;.
\label{27}
\eea
The ten-dimensional IIB theory is recovered from ${\rm E}_{6(6)}$ exceptional field theory upon solving the associated 
section condition by restricting the coordinate dependence of
all fields to 5 coordinates within the $(15,1)$ (transforming as a vector under the maximal ${\rm GL}(5)$ subgroup).
Specifically, with (\ref{27}), we decompose coordinates as
\bea
Y^{M}&\longrightarrow& \left\{Y^{[AB]},\; Y_{A\alpha} \right\}\;,\qquad
\mbox{with}\quad A=0, \dots, 5\;,\quad \alpha=1,2
\;,
\eea
and impose
\bea
\partial_{ij} \Phi &=& 0\;,\quad \partial^{0\alpha} \Phi ~=~ 0\;,\quad \partial^{i\alpha} \Phi ~=~ 0\;,
\label{sol_27}
\eea
for $i=1, \dots, 5$\,.
Comparing to (\ref{section_SL}), we observe that the ${\rm SL}(6)$ twist matrix constructed above
is compatible with this solution of the section condition. Upon the embedding (\ref{662}), (\ref{27}), it
gives rise to an ${\rm E}_{6(6)}$ twist matrix $U_M{}^N$ 
\bea
U_M{}^N &=& 
 \left(
\begin{array}{c:c}
U_{[AB]}{}^{[CD]}  & 0 
\\ \hdashline
0 & \delta_\gamma^\alpha\,(U^{-1})_{C}{}^{A}   
\end{array}
\right)
\;,
\label{U6}
\eea
satisfying (\ref{sol_27}).
As a consequence, the
generalized Scherk-Schwarz ansatz (\ref{SSeAB}) for the full \eft
\bea
 e_\mu{}^\alpha(x,Y) &=& \rho^{-1}(Y)\,e_\mu{}^\alpha(x)
 \;,\nonumber\\
{\cal M}_{MN}(x,Y) &=& U_{M}{}^{P}(Y)\,U_{N}{}^{Q}(Y)\,{M}_{PQ}(x)\;,
\nonumber\\  {\cal A}_{\mu}{}^{M}(x,Y) &=& A_{\mu}{}^{N}(x)(U^{-1})_{N}{}^{M}(Y)\,\rho^{-1}(Y)\;, 
  \nonumber\\
  {\cal B}_{\mu\nu\,M}(x,Y) &=& \,\rho^{-2}(Y)\, U_{M}{}^{P}(Y)\,B_{\mu\nu\,P}(x)\;,
 \eea
describes consistent truncations from IIB supergravity to $D=5$ theories, provided the 
twist matrix (\ref{U6}) solves the full set
of consistency conditions (\ref{consistent_general}). 

Let us thus consider the matrix (\ref{U6}) built from the 
${\rm SL}(6)$ matrix $U_A{}^B$ from (\ref{Upqr}), (\ref{beta}), which in turn solves equations
(\ref{con1})--(\ref{con3}) for general $\eta_{AB}$ characterised by three integers $p+q+r=6$. 
The first consistency equation (\ref{consistent_yet}) for the ${\rm E}_{6(6)}$ twist matrix (\ref{U6}) reads
\bea
\rho^{-1}\,\left[ {\cal X}_M{}^\bfa \right]_{({\bf 351})} &\stackrel{!}{=}&
 {\rm const}
 \;,
 \label{con61}
\eea
with ${\cal X}_M{}^\bfa$ defined in (\ref{XXX}). 
It follows from the form of the matrix (\ref{U6}) together with (\ref{sol_27}) that the only non-vanishing 
components of ${\cal X}_{M}{}^\bfa$ from (\ref{XXX}) are its components 
${\cal X}_{[AB]}{}^\bfa$ when $\bfa$ takes values in the ${\rm SL}(6)$ subgroup of ${\rm E}_{6(6)}$.
These are nothing but the ${\cal X}_{AB,C}{}^D$ defined directly in terms of $U_A{}^B$ in (\ref{XX}) above,
and moreover singlets under the ${\rm SL}(2)$ of (\ref{662}).
For equation (\ref{con61}) this means that under decomposition w.r.t.\ ${\rm SL}(6)$ its only non-vanishing
components are the ${\rm SL}(2)$ singlets in the branching of ${\bf 351}\longrightarrow (21,1)+(105,1)+\dots$,
reducing to
\bea
\rho^{-1}\,\left[ {\cal X}_M{}^\bfa \right]_{(21)} &\stackrel{!}{=}&
 {\rm const}
 \;,\qquad
 \rho^{-1}\,\left[ {\cal X}_M{}^\bfa \right]_{(105)} ~\stackrel{!}{=}~
 {\rm const}
 \;.
 \label{con62}
\eea
Comparing to the general discussion around (\ref{conNcon}), (\ref{conNnot}),
these equations are precisely ensured by the properties of the ${\rm SL}(6)$ twist matrix $U_A{}^B$ 
constructed above, as a consequence of (\ref{consistent_SL}).
Specifically, we find that
\bea
\rho^{-1}\,\left[ {\cal X}_M{}^\bfa \right]_{(AB)} &=&
 \eta_{AB}
 \;,\qquad
 \rho^{-1}\,\left[ {\cal X}_M{}^\bfa \right]_{(105)} ~=~
 0
 \;,
 \label{con63}
\eea
and conclude that the first consistency equation in (\ref{consistent_general}) 
is solved by (\ref{U6}) with the density factor $\rho$ given by (\ref{rhopqr}) above.
It remains to study the second equation from~(\ref{consistent_general}). Again, the structure of the matrix (\ref{U6})
and its coordinate dependence (\ref{sol_27}) imply that the l.h.s.\ of this equation reduces to
\bea
  \partial_{CD} (U^{-1})_{AB}{}^{CD}  
- (D-1)\,\rho^{-1}\partial_{CD} \rho  \, (U^{-1})_{AB}{}^{CD}
\;,
\eea
which vanishes for the ${\rm SL}(6)$ twist matrix $U_A{}^B$ due to its property (\ref{con2}),
by virtue of the integer relation (\ref{ddd}), which holds for the present case $(D,n)=(5,6)$\,.

We conclude, that the twist matrix (\ref{U6}) with $U_A{}^B$ given by  
(\ref{Upqr}), (\ref{beta}) above, together with the density factor $\rho_{p,q,r}$ from (\ref{rhopqr}) 
satisfies both the section constraints
(as a subclass of the general IIB solution (\ref{sol_27})), and the consistency
equations (\ref{consistent_general}).
Via (\ref{con63}) it corresponds to
an embedding tensor in the $21$ of ${\rm SL}(6)$
parametrized by the diagonal matrix $\eta_{AB}$ from (\ref{YinCSO}).
The generalized Scherk-Schwarz ansatz thus describes the consistent truncation from 
$D=10$ IIB supergravity to a maximal $D=5$ gauged supergravity with 
gauge group ${\rm CSO}(p,q,r)$\,.

\subsubsection{${\rm E}_{7(7)}$}

This case works in complete analogy to  ${\rm E}_{6(6)}$. We have reviewed the 
${\rm E}_{7(7)}$ exceptional field theory in section~\ref{sec:eftrev} above.
The relevant subgroup is ${\rm SL}(8)$ embedded into ${\rm E}_{7(7)}$
such that the 56 internal coordinates decompose as
\bea
{\bf 56} &\longrightarrow& 
28 + 28'\;,\qquad
Y^{M}~\longrightarrow~ \left\{Y^{[AB]},\; Y_{[AB]} \right\}
\;.
\label{56}
\eea
The full $D=11$ theory is recovered from ${\rm E}_{7(7)}$ exceptional field theory upon solving the associated 
section condition by restricting the coordinate dependence of
all fields as
\bea
\partial_{ij} \Phi &=& 0\;,\qquad \partial^{AB} \Phi ~=~ 0\;,
\label{sol_56}
\eea
for $i=1, \dots, 7$, $A=0, \dots, 7$\,.
Comparing to (\ref{section_SL}), we observe that the ${\rm SL}(8)$ twist matrix constructed above
is compatible with this solution of the section condition. Upon the embedding (\ref{56}), it
gives rise to an ${\rm E}_{7(7)}$ twist matrix $U_M{}^N$ 
\bea
U_M{}^N &=& 
 \left(
\begin{array}{c:c}
U_{[AB]}{}^{[CD]}  & 0 
\\ \hdashline
0 & (U^{-1})_{[CD]}{}^{[AB]}   
\end{array}
\right)
\;,
\label{U7}
\eea
satisfying (\ref{sol_56}).
As a consequence, the
generalized Scherk-Schwarz ansatz (\ref{SSeAB}) for the full \eft
describes consistent truncations from $D=11$ supergravity to $D=4$ theories, provided the 
twist matrix (\ref{U7}) solves the full set
of consistency conditions (\ref{consistent_twist}). 

The first of these equations written in the form (\ref{consistent_yet}) reads
\bea
\rho^{-1}\,\left[ {\cal X}_M{}^\bfa \right]_{({\bf 912})} &\stackrel{!}{=}&
 {\rm const}
 \;,
 \label{con71}
\eea
with ${\cal X}_M{}^\bfa$ defined in (\ref{XXX}). 
It follows from the form of the matrix (\ref{U7}) together with (\ref{sol_56}) that the only non-vanishing 
components of ${\cal X}_{M}{}^\bfa$ from (\ref{XXX}) are its components 
${\cal X}_{[AB]}{}^\bfa$ when $\bfa$ takes values in the ${\rm SL}(8)$ subgroup of ${\rm E}_{7(7)}$.
As for the ${\rm E}_{6(6)}$ case, these are nothing but the ${\cal X}_{AB,C}{}^D$ defined directly in terms 
of $U_A{}^B$ in (\ref{XX}) above.
For equation (\ref{con61}) this means that under decomposition w.r.t.\ ${\rm SL}(8)$ its only non-vanishing
components are given by
\bea
\rho^{-1}\,\left[ {\cal X}_M{}^\bfa \right]_{(36)} &\stackrel{!}{=}&
 {\rm const}
 \;,\qquad
 \rho^{-1}\,\left[ {\cal X}_M{}^\bfa \right]_{(420)} ~\stackrel{!}{=}~
 {\rm const}
 \;.
 \label{con72}
\eea
Comparing to the general discussion around (\ref{conNcon}), (\ref{conNnot}),
these equations are precisely ensured by the properties of the ${\rm SL}(8)$ twist matrix $U_A{}^B$ 
constructed above, as a consequence of (\ref{consistent_SL}).
Specifically, we find that
\bea
\rho^{-1}\,\left[ {\cal X}_M{}^\bfa \right]_{(AB)} &=&
 \eta_{AB}
 \;,\qquad
 \rho^{-1}\,\left[ {\cal X}_M{}^\bfa \right]_{(420)} ~=~
 0
 \;,
 \label{con73}
\eea
and conclude that the first consistency equation in (\ref{consistent_twist}) 
is solved by (\ref{U6}) with the density factor $\rho$ given by (\ref{rhopqr}) above.
It remains to study the second equation. Again, the structure of the matrix (\ref{U7})
and its coordinate dependence (\ref{sol_56}) imply that the l.h.s.\ of this equation reduces to
\bea
  \partial_{CD} (U^{-1})_{AB}{}^{CD}  
- 3\,\rho^{-1}\partial_{CD} \rho  \, (U^{-1})_{AB}{}^{CD}
\;,
\eea
which vanishes for the ${\rm SL}(8)$ twist matrix $U_A{}^B$ due to its property (\ref{con2}),
by virtue of the integer relation (\ref{ddd}), which holds for the present case $(D,n)=(4,8)$\,.
In full analogy to the ${\rm E}_{6(6)}$ case we find that with the ansatz (\ref{U7}) 
for the ${\rm E}_{7(7)}$ matrix $U_M{}^N$, 
all non-vanishing parts of the consistency equations (\ref{consistent_general}) precisely reduce to the 
corresponding equations (\ref{consistent_SL}) for the ${\rm SL}(n)$ matrix $U_A{}^B$.

We conclude, that the twist matrix (\ref{U7}) with $U_A{}^B$ given by  
(\ref{Upqr}), (\ref{beta}) above, together with the density factor $\rho_{p,q,r}$ from (\ref{rhopqr}) 
satisfies both the section constraints
(as a subclass of the general $D=11$ solution (\ref{sol_56})), and the consistency
equations (\ref{consistent_twist}).
Via (\ref{con73}) it corresponds to
an embedding tensor in the ${\bf 36}$ of ${\rm SL}(8)$
parametrized by the diagonal matrix $\eta_{AB}$ from (\ref{YinCSO}).
The generalized Scherk-Schwarz ansatz thus describes the consistent truncation from 
$D=11$ supergravity to a maximal $D=4$ gauged supergravity with 
gauge group ${\rm CSO}(p,q,r)$\,.

\subsection{The induced space-time metric}
\label{subsec:induced}

In the above, we have constructed twist matrices as solutions of the consistency equations (\ref{consistent_general})
which define consistent truncations of the higher-dimensional $D=11$ and IIB supergravity down to 
$D=4, 5, 7$ maximal supergravity with gauge group ${\rm CSO}(p,q,r)$\,.
While consistency of the truncation follows from the general structure of the ansatz and the covariant
formulation of exceptional field theory,
for physical applications one will typically be interested in the explicit embedding of the lower-dimensional fields 
into $D=11$ and IIB supergravity in their original form. 
The translation of the very compact ansatz (\ref{SSeAB}) into the original fields of higher-dimensional supergravity
thus requires the explicit dictionary between the fields of the exceptional field theory and the original supergravities
\cite{Hohm:2013vpa,Hohm:2013uia}.

As an example, let us work out the form of the internal background metric to which the above compactifications correspond.
The internal components of the higher-dimensional metric sit among the components of the scalar matrix 
${\cal M}_{MN}=({\cal V}{\cal V}^T)_{MN}$, built from a group-valued vielbein ${\cal V}$ that carries the higher-dimensional
components according to the decomposition of the Lie algebra w.r.t.\ to a proper grading, c.f.\ (\ref{V56}).
See~\cite{Cremmer:1997ct} for the general structure of these parametrizations, and~\cite{Berman:2011jh} for some explicit matrices.

\begin{figure}[tb]
 \centering
\includegraphics[scale=.8]{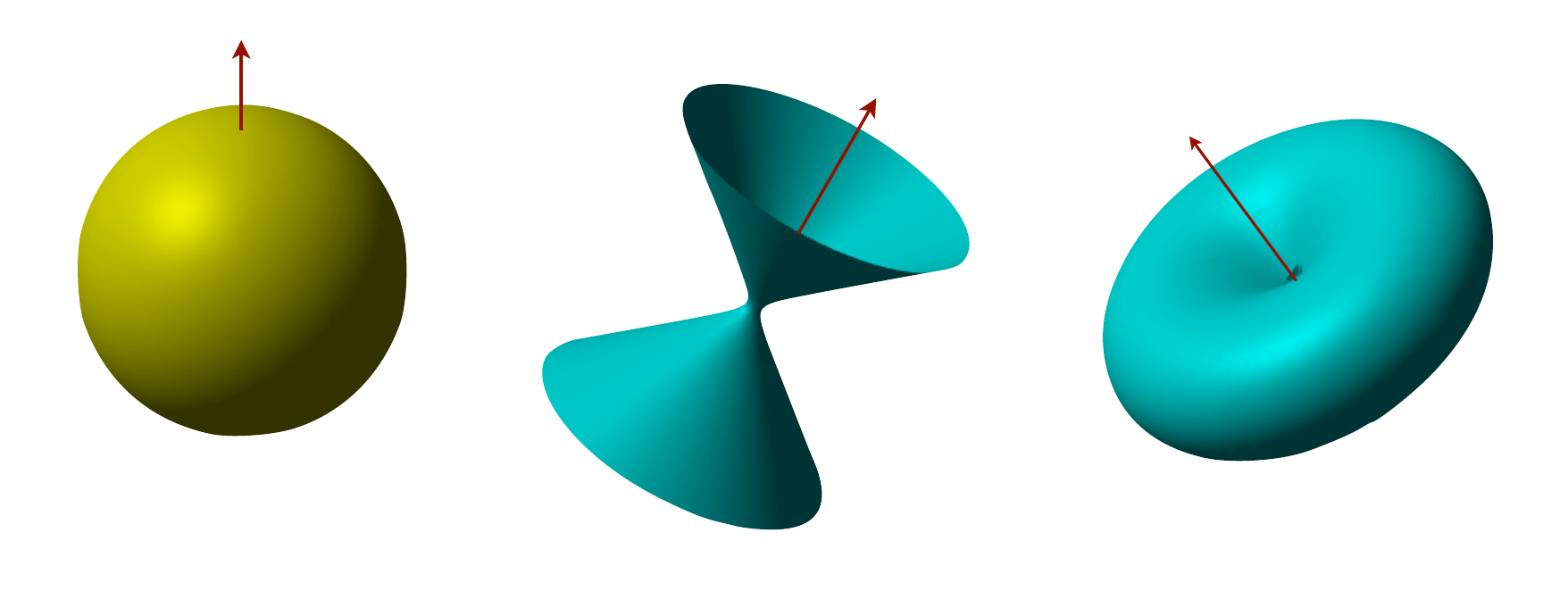}
\caption{{\small Sphere, hyperboloid and the warped hyperboloid~(\ref{internalG}) in a $2d$ projection,
i.e.\ for $(p,q,r)=(2,1,0)$.
}}
\label{hyper}
\end{figure}

As a general feature of the theories with ${\rm SL}(n)$ embedding according to (\ref{SLinE1}), (\ref{SLinE2}), 
the generalized metric ${\cal M}^{MN}$ decomposes into blocks
\bea
{\cal M}^{MN} &=& 
 \left(
\begin{array}{c:c:c}
{\cal M}^{0i,0j}  & {\cal M}^{0i,jk} & \;\cdots
\\ \hdashline
{\cal M}^{ij,0k}  & {\cal M}^{ij,kl} & \;\cdots
\\ \hdashline
\vdots & \vdots & \;\ddots
\end{array}
\right)
\;,
\label{Mmm}
\eea
and the higher-dimensional internal metric $g_{ij}$
can be read off from the upper left block as
\bea
{\cal M}^{i0,j0} &=& g^{(4-n)/n}\,g^{ij}
\;.
\label{Ug}
\eea
The power of $g$ on the r.h.s.\ is fixed by the fact that generalized diffeomorphisms on ${\cal M}^{MN}$
translate into standard diffeomorphisms for $g^{ij}$\,.\footnote{
A short calculation shows compatibility with the explicit result (5.25) of~\cite{Hohm:2013vpa} for ${\rm E}_{6(6)}$\,.}
For the moment, we are just interested in the higher-dimensional metric at the 'origin' of the truncation, i.e.\ at the 
point where all lower-dimensional scalar fields vanish. According to (\ref{genSS}), at this point, ${\cal M}_{MN}$ is given by
\bea
{\cal M}^{MN} &=& (U^{-1})_P{}^M\,(U^{-1})_P{}^N
\;,
\label{MUU}
\eea
in terms of the twist matrix $U$\,. Since $U$ is embedded in the subgroup ${\rm SL}(n)$, c.f.\ (\ref{U6}), (\ref{U7}),
the relevant block (\ref{Ug}) of ${\cal M}$ can simply be expressed as
\bea
{\cal M}^{i0,j0} &=& 
\frac12\left(
m^{ij}m^{00}-m^{i0}m^{j0}
\right)
\;,
\eea
for $m^{AB} \equiv (U^{-1})_C{}^{A}\,(U^{-1})_C{}^{B}$\,.
With the explicit form (\ref{Upqr}) of the twist matrices constructed in section~\ref{subsec:sphere}
for gauge group ${\rm CSO}(p,q,r)$,
we can thus work out the internal metric, and find after some calculation
\bea
ds^2 &=& g_{ij}\, dy^i\,dy^j \nonumber\\
&=& (1+u-v)^{-2/(p+q+r-2)} \left(
dy^z dy^z + dy^a dy^b \left(\delta^{ab} +\frac{\eta_{ac} \eta_{bd} {y}^c {y}^d}{1-v} \right)\right)
\;,
\label{internalG}
\eea
with the split of coordinates $y^i=\{y^a, y^u\}$, $a=1, \dots, p+q-1$, and $z=p+q, \dots, r$\,,
and the combinations $u\equiv y^a y^a$, $v\equiv y^a\eta_{ab}y^b$\,.
This space is conform to the direct product
\bea
H^{p,q} \times \mathbb{R}^r
\;,
\eea
of $r$ flat directions and the hyperboloid $H^{p,q}$\,. The latter is defined by the embedding surface
\bea
y^A \,\eta_{AB}\, y^B &=& 1
\;,\qquad A=0, \dots, p+q-1
\;,
\eea
with $\eta_{AB}$ from (\ref{YinCSO}), within a $(p+q)$ dimensional Euclidean space.
For $r=q=0$, the metric (\ref{internalG}) is the round sphere (the pre-factor becomes constant: $1+u-v\rightarrow1$).
For $p+q+r=8$, this is precisely the metric proposed by Hull and Warner in~\cite{Hull:1988jw}
with the warp factor deforming the hyperboloid geometry, see figure~\ref{hyper}.
This is 
the higher-dimensional background inducing the ${\rm CSO}(p,q,r)$ gauged supergravities 
in $D=4$ dimensions.
Along the very same lines, the higher-dimensional metric can be computed for arbitrary values of the lower-dimensional
scalar fields, i.e.\ for arbitrary values of the matrix $M_{PQ}(x)$, in which case (\ref{MUU}) is replaced by the full
Scherk-Schwarz ansatz (\ref{genSS}).
The uplift of all the $D$-dimensional fields, i.e.\ all the non-linear reduction ans\"atze follow straightforwardly 
(although by somewhat lengthy calculation) from combining this ansatz
with the dictionary of the full exceptional field theory to higher-dimensional supergravity, which is independent of the 
particular form of the twist matrix.

We stress once more, that the metric (\ref{internalG}) is in general not part of a solution of the 
higher-dimensional field equations. This simply translates into the fact that the corresponding lower-dimensional supergravity
in general does not possess a solution with all scalar fields vanishing. 
Indeed, it was shown in \cite{Hull:1988jw} for the ${\rm SO}(p,8-p)$ supergravities that the metric (\ref{internalG})
is compatible with a generalized Freund-Rubin ansatz only for the values $(p,q)=(8,0)$, $(p,q)=(4,4)$, and $(p,q)=(5,3)$,
which precisely correspond to the 
gauged supergravities admitting critical points at the origin of the scalar potential.
Independently of this property concerning the ground state, in all cases the generalized Scherk-Schwarz ansatz (\ref{genSS}), (\ref{SSeAB}),
continues to describe the consistent truncation of the higher-dimensional supergravity to the field content and the dynamics of a
lower-dimensional maximally supersymmetric supergravity.

\section{Conclusions}
\label{sec:conclusions}

In this paper we have shown how the consistency of a large class of Kaluza-Klein truncations 
can be proved using exceptional field theory. The Kaluza-Klein ansaetze are given in terms of 
twist matrices $U_{M}{}^{N}$ taking values in the duality group E$_{d(d)}$ subject to the consistency
equations~(\ref{consistent_general}). The resulting effective  gauged supergravities emerge naturally in
the embedding tensor formalism, upon identifying the embedding tensor with a particular projection of $U^{-1}\partial U$. 
Such generalized Scherk-Schwarz reductions have been considered before
in various studies for truncations of the full exceptional field theory.
Here, we have given the Scherk-Schwarz ansatz for the full field content of exceptional field theory,
including fermions, $p$-forms, and the constrained compensating tensor gauge fields.
We have then shown that the ansatz reproduces the complete field equations of the lower-dimensional gauged supergravity.
This is necessary in order to relate to the full untruncated $D=11$ or type IIB supergravity.  
Secondly, we explicitly constructed the twist matrices for various compactifications, 
including sphere compactifications such as 
AdS$_5\times {\rm S}^5$, and new examples such as hyperboloids $H^{p,q}$. In 
contrast to ordinary Scherk-Schwarz compactifications where the existence of consistent twist matrices
is guaranteed by the underlying Lie algebraic structure of the deformation parameters, to our knowledge there is no analogue of 
such an existence proof for solutions of the generalized consistency equations (\ref{consistent_general}) 
with general embedding tensor $\Theta_M{}^{\bfa}$\,.
In this regard, the construction of explicit examples is a crucial step towards a more
systematic understanding of the underlying structures.

Given the explicit form of the twist matrices, any solution of 
gauged supergravity is embedded into the higher-dimensional exceptional field theory 
via the relations (\ref{genSS_intro})--(\ref{BM_intro}), and thereby further into  $D=11$ or type IIB supergravity. 
The explicit embedding formulas into the original $D=11$ or type IIB supergravity require
the dictionary relating the exceptional field theory fields to the original formulation of these theories.
It should be stressed that this dictionary is completely independent of 
the particular compactification or twist matrix and can be fixed, for instance,
by matching the gauge symmetries on both sides. 
Together this allows in particular to lift the known AdS solutions~\cite{Fischbacher:2011jx} 
of ${\rm SO}(8)$ gauged supergravity
to eleven dimensions, but also the large class of dS and domain wall solutions found in the 
non compact and non-semisimple four-dimensional gaugings~\cite{Hull:1984rt,Ahn:2001by,Dall'Agata:2012sx}.

It will be interesting to explore the possible generalizations of the presented construction.
Our construction of twist matrices was based on the maximal embedding (\ref{SLinE}), (\ref{SLinE1})
of an ${\rm SL}(n)$ group into the corresponding exceptional group. For the ${\rm SO}(p,q)$ case,
as a consequence of the second equation in (\ref{consistent_general}), this led
to the severe constraint (\ref{ddd}), restricting the construction to the three principal cases 
$(D,n)=(7,5), (5,6), (4,8)$. In the ${\rm CSO}(p,q,r)$ case on the other hand, 
the structure of the twist matrix (\ref{Upqr}) suggests that there is more freedom due to the possible 
introduction of the factor $\beta$, preserving the ${\rm CSO}(p,q,r)$ structure.
As a result, the construction will still go through without being constrained to (\ref{dddpqr}),
i.e.\ in particular for other values of $D$, as given in (\ref{beta0}). This will be interesting to explore.
Similarly, the construction should allow for more solutions, when the embedding 
(\ref{SLinE1}) of ${\rm SL}(n)$ is not maximal but leaves additional abelian factors. 
This is the case for the $D=6$, ${\rm G}={\rm SO}(5,5)$ \eft with ${\rm SL}(5)$ embedded
via the intermediate ${\rm GL}(5)$\,. The additional ${\rm GL}(1)$ factor allows for an additional
parameter in the twist matrix, that can be tuned such as to solve the second equation 
in (\ref{consistent_general}) without reverting to (\ref{ddd}). This case should thus include the consistent $S^4$
reduction of IIA supergravity~\cite{Cvetic:2000ah}.

A different class of twist matrices should correspond to solutions of (\ref{consistent_SL})
in which (\ref{ZT}) is relaxed to non-vanishing $Z_{ABC}{}^D$\,.
E.g.\ $D=7$ maximal supergravity possesses a gauging with $\eta_{AB}=0$ and compact ${\rm SO}(4)$
gauge group \cite{Samtleben:2005bp} which is conjectured to correspond to a consistent sphere truncation
of IIB supergravity. It would be interesting to work out the corresponding twist matrix and to explore
its generalization to arbitrary values of $n$. It should of course also be possible to find twist matrices that
generate a non-vanishing trombone parameter $\vartheta_M$\,. More generally, one may try to 
generalize the above construction by replacing (\ref{SLinE1}) to an embedding of coordinates via other classes of subgroups.
The method should also extend to non-maximal theories such as the AdS$_3\times S^3$ reduction
from six dimensions \cite{Deger:1998nm,Deger:2014ofa}.

Another extension of our results that may eventually become feasible is 
the inclusion of higher-derivative $\alpha'$ or M-theory corrections. 
Indeed, in double field theory there was progress recently 
of how to include $\alpha'$ corrections \cite{Hohm:2013jaa,Bedoya:2014pma,Hohm:2014eba,Hohm:2014xsa}, 
see also \cite{Anderson:2014xha,Coimbra:2014qaa,delaOssa:2014msa}.
In particular, the results of  \cite{Hohm:2013jaa} provide an exactly O$(d,d)$ invariant description 
of a subsector of heterotic string theory to all orders in $\alpha'$. 
If a generalization to exceptional field theory exists one may hope for 
consistent Kaluza-Klein embeddings that not only lead to exact solutions of the 
higher-dimensional field equations but also to solutions that are exact in $\alpha'$.

Finally, let us stress that throughout  this paper we assumed the strong form of the section 
constraints~(\ref{sectioncondition}). Thereby the twist matrices we construct as solutions of (\ref{consistent_general})
describe consistent truncations within the original $D=11$ and IIB supergravity.
It is intriguing, however, that the match with lower-dimensional gauged supergravity,  
upon reduction by the Scherk-Schwarz ansatz, does not explicitly use the section constraint
(provided the initial scalar potential is written in an appropriate form), 
as observed in \cite{Berman:2012uy,Aldazabal:2013mya}
for the internal sector and shown here for the full theory. 
Formally this allows to reproduce all gauged supergravities, 
and it is intriguing to speculate about their higher-dimensional embedding 
upon possible relaxation of the section constraints that would 
define a genuine extension of the original supergravity theories. 
For the moment it is probably fair to say that our understanding of a consistent extension of the framework is 
still limited. In this context it would be interesting to obtain explicit examples of twist matrices that satisfy all consistency conditions
(\ref{consistent_general}), but violate the section constraints~(\ref{sectioncondition}) which may give a hint as
to how to consistently relax these constraints in exceptional field theory.
We hope that our treatment of sphere and hyperboloid compactifications may help clarify these matters.

\subsection*{Acknowledgements}
We would like to thank Hadi and Mahdi Godazgar, Falk Hassler, Dieter L\"ust, 
Diego Marques  and Mario Trigiante for interesting discussions. 
This work is supported by the U.S. Department of Energy (DoE) under the cooperative research agreement DE-FG02-05ER41360. The work of O.H. is supported by a DFG Heisenberg fellowship.

\section*{Appendix}

\begin{appendix}

\section{Projection of the spin connection}

Consider a general tensor $W_{MN}{}^K = W_{M}{}^{\bfa}\,(t_\bfa)_N{}^K$,
living in the full tensor product 
\bea
{\bf 56}\otimes {\bf 133}={\bf 56}\oplus{\bf 912}\oplus{\bf 6480}
\;.
\label{56x133}
\eea
In flattened $SU(8)$ indices, it takes the matrix form
\bea
W_{ij} &=&
 \begin{pmatrix}
 - \frac23\delta_{[k}{}^{[p}\, W^{q]}{}_{l]ij} & 
    {1\over 24}\epsilon_{klrstuvw} \, W^{tuvw}{}_{ij} \cr
    \noalign{\vskip5mm}
      W^{mnpq}{}_{ij} &
             \frac23 \delta_{[r}{}^{[m} \,W^{n]}{}_{s]ij}\cr
             \end{pmatrix}
             \;,
\eea
and complex conjugate.
The diagonal and off-diagonal blocks are parametrized by 
$W^{i}{}_{j[kl]}$, $W^{[tuvw]}{}_{ij}$ with $W^i{}_{ikl}=0$\,.
In general, the diagonal and off-diagonal blocks carry the ${\rm SU}(8)$ representations
\bea
W^i{}_{jkl} &:& {\bf 28} \oplus {\bf 36} \oplus {\bf 420} \oplus {\bf 1280} \;,\nonumber\\
W^{tuvw}{}_{ij} &:& \overline{{\bf 28}}\oplus \overline{{\bf 420}} \oplus {\bf 1512}
\;.
\label{WW}
\eea
Together with their complex conjugates they fill the three irreducible
${\rm E}_{7(7)}$ representation~ (\ref{56x133}). The ${\bf 36}\oplus\overline{{\bf 36}}$ sit in the ${\bf 912}$, 
on the other hand there are two
copies of the ${\bf 420}\oplus\overline{\bf 420}$ sitting in the ${\bf 912}$ and the ${\bf 6480}$,
respectively. In order to disentangle the
different representations, it is useful to recollect the transformation of $W_i{}^{jkl}$,
$W^{ijkl}{}_{mn}$ under the 70 ${\rm E}_{7(7)}/{\rm SU}(8)$ generators,
which mix these fields as follows~\cite{deWit:2007mt}
\begin{eqnarray}
  \label{WE7}
  \delta W_i{}^{jkl} &=&  2\, \Sigma^{jmnp}\, W_{imnp}{}^{kl} -\ft14\,\d_i{}^j\, \Sigma^{mnpq}\, 
  W_{mnpq}{}^{kl} + \Sigma^{klmn}\, W^j{}_{imn} \,, \nn\\
  \delta W_{ijkl}{}^{mn} &=& {}-\ft43 \Sigma_{p[ijk}^{~}\, W_{l]}{}^{pmn}
  -\ft1{24} \varepsilon_{ijklpqrs}\, \Sigma^{mntu}_{~}\, W^{pqrs}{}_{tu}
  \,.   
\end{eqnarray}
Iterating this transformation, we can compute the action of the 
${\rm E}_{7(7)}/{\rm SU}(8)$ Casimir $\Delta\equiv\delta_{ijkl} \delta^{ijkl}$
\bea
\Delta W_i{}^{jkl} &=& 
\frac{35}{6}\,W_i{}^{jkl} -\frac{5}{12}\,W_m{}^{mkl} \delta_i{}^j
-\frac12\,T^{klmn}{}_{mn}\,\delta_i{}^j
-2\,\delta_i{}^{[k}\,W^{l]jmn}{}_{mn} + 2\,W^{jklm}{}_{im}
\;,\nonumber\\
\Delta W^{klmn}{}_{ij} &=& 
\frac{11}6\,W^{klmn}{}_{ij} + \frac43\,
W_{[i}{}^{[klm} \delta_{j]}{}^{n]}-\frac43\,
\delta_{ij}{}^{[mn}\,W_{p}{}^{kl]p}-\frac13\,
W_{p}{}^{p[kl}\,\delta_{ij}{}^{mn]}
\;,
\label{Casimir}
\eea
whose different eigenvalues allow to identify the ${\rm E}_{7(7)}$ origin
of the various ${\rm SU}(8)$ blocks. E.g.\ parametrizing the two ${\bf 420}$ representations as
\bea
W_i{}^{jkl} &=&A_i{}^{jkl}+\cdots\;,\qquad
W^{klmn}{}_{ij} ~=~ \delta_{[i}{}^{[k}\,B_{j]}{}^{lmn]}+\cdots
\;,
\label{WAB}
\eea
with traceless $A_{i}{}^{[jkl]}$, $B_{i}{}^{[jkl]}$, 
the action (\ref{Casimir}) diagonalises on the combinations
\bea
4\,A_i{}^{jkl} &=& -3\,B_i{}^{jkl}
\;,\qquad\mbox{with eigenvalue}\;\;\frac92
\;,\nonumber\\
A_i{}^{jkl} &=& B_i{}^{jkl}
\;,\qquad\mbox{with eigenvalue}\;\;\frac{41}6
\;.
\label{comb12}
\eea
The first option corresponds to the {\bf 420} from the ${\bf 912}$, cf.~(3.28) of \cite{deWit:2007mt}, the other one
thus is the {\bf 420} from the ${\bf 6480}$\,. We conclude, that the projection of (\ref{WAB}) onto the {\bf 420} from the
${\bf 912}$ is given by
\bea
\frac37 \left(A_i{}^{jkl} - B_i{}^{jkl}\right)
\;,
\eea
such that consistently the second combination of (\ref{comb12}) is projected to zero, and the first one
is projected to itself.

Putting everything together, we learn that taking the original spin connection $Q_{AB}{}^C$ living in the ${\rm SU}(8)$
of the form (schematically)
\bea
Q_{ij} &=& \left(
\begin{array}{c:c}
A_{420}+\dots & 0 
\\ \hdashline
0 & A_{420}+\dots
\end{array}
\right)
\;,
\label{Q1}
\eea
i.e.\ with $B_i{}^{jkl}=0$,
its projection onto the ${\bf 912}$ maps this into a matrix of the form
\bea
[Q_{ij}]_{(\bf 912)} &=& \left(
\begin{array}{c:c}
\frac37 A_{420}+\dots & \frac37 A_{420}+\dots 
\\ \hdashline
\frac37 A_{420}+\dots & \frac37 A_{420}+\dots
\end{array}
\right)
\;.
\label{Q2}
\eea
This is the form of the matrix which via (\ref{SSTrel}) we identify with the $T$-tensor 
of gauged supergravity \cite{deWit:2007mt}, parametrized by $A_1$, $A_2$ as in (\ref{A1A2DEF})\,.
Comparing (\ref{Q1}), (\ref{Q2}) we see that upon projection, a relative factor of $\frac37$ has to be taken into
account in the ${\bf 420}$ part $A_{2\,i}{}^{jkl}$, while the ${{\bf 36}}$ part $A_{1}^{ij}$
(which is unique in (\ref{WW})) remains unchanged.

\end{appendix}


\providecommand{\href}[2]{#2}\begingroup\raggedright\endgroup

\end{document}